\documentclass[letterpaper]{JHEP3} % 10pt is ignored!
\usepackage{epsfig,array,cite,latexsym,amsmath}%,  showkeys, oldgerm}
\preprint{BUHEP-06-04}
\newcommand\eqn[1]{\label{eq:#1}} 
\newcommand\Eq[1]{Eq.~\eqref{eq:#1}} 
\newcommand\half{{\textstyle{\frac{1}{2}}}} 
\newcommand\fourth{{\textstyle{\frac{1}{4}}}} 
\newcommand\fourthi{{\textstyle{\frac{i}{4}}}}
\newcommand\eight{{\textstyle{\frac{1}{8}}}}

\newcommand\wchi{\widetilde{\chi}}
\newcommand\wQ{\widetilde{Q}}
\newcommand{\beq}{\begin{eqnarray}}
\newcommand{\eeq}{\end{eqnarray}}

\newcommand{\CD}{{\cal D}}
\newcommand{\CE}{{\cal E}}  
\newcommand{\CF}{{\cal F}}

\newcommand{\CN}{{\cal N}}

\newcommand{\CQ}{{\cal Q}}

\newcommand{\CL}{{\cal L}}
\newcommand{\bfn}{{\bf n}}
\newcommand{\bfr}{{\bf r}}

\newcommand{\bfe}{{\bf  e}}

\DeclareMathOperator{\Tr}{Tr}

\newcommand{\sla}[1]%
        {\kern .25em\raise.18ex\hbox{$/$}\kern-.55em #1}

\newcommand{\mybar}[1]%
        {\kern 0.6pt\overline{\kern -0.6pt#1\kern -0.6pt}\kern 0.6pt}
\newcommand{\dig}{\kern-1.5pt \raisebox{.9ex}{$\cdot$}  \kern1.5pt
  \raisebox{0ex}{${\mathbf\cdot}$}\kern1.5pt \raisebox{-.9ex}{$\cdot$}} 
\newcommand{\digb}{\kern-1.5pt \raisebox{.75ex}{$\cdot$}  \kern1.5pt
  \raisebox{0ex}{${\mathbf\cdot}$}\kern1.5pt \raisebox{-.75ex}{$\cdot$}} 
\newcommand{\digc}{\kern-1.5pt \raisebox{1.05ex}{$\cdot$}  \kern1.5pt
  \raisebox{0ex}{${\mathbf\cdot}$}\kern1.5pt \raisebox{-1.05ex}{$\cdot$}} 
\newcommand\fverb{\setbox\pippobox=\hbox\bgroup\verb}
\newcommand\fverbdo{\egroup\medskip\noindent%
                        \fbox{\unhbox\pippobox}\ }
\newcommand\fverbit{\egroup\item[\fbox{\unhbox\pippobox}]}
\newbox\pippobox

\title{ Twisted   Supersymmetric Gauge  Theories 
and Orbifold  Lattices }
\author
    {%
     Mithat \"Unsal
    \\Department of Physics
    \\Boston University 
    \\590 Commonwealth Ave, Boston, MA 02215
    \\Email: 
    \parbox[t]{2in}{\email {unsal@buphy.bu.edu} }
    }%
\abstract{We examine  the relation  between    twisted versions of the  
extended  supersymmetric gauge theories and  
supersymmetric orbifold lattices.  
In particular, for the   $\CN=4$  SYM in $d=4$, we show that the  continuum 
limit of orbifold lattice reproduces the twist introduced by 
Marcus, and the examples at lower dimensions are usually 
Blau-Thompson type.  
The orbifold  lattice point group symmetry is  
a subgroup of the twisted Lorentz group,  and the exact supersymmetry of the 
lattice is indeed the nilpotent scalar supersymmetry  of the twisted 
versions. We also introduce   twisting in terms of spin groups 
of  finite point subgroups of $R$-symmetry and spacetime symmetry. 
}

\begin{document} 

\section{Introduction} 
This paper is devoted to the study of the  relation between  
supersymmetric  orbifold lattices and  twisted versions  of  extended 
supersymmetric  gauge theories.   
This turns out to be useful in many respects. The  viewpoint of this paper 
explains 
many oddities of orbifold lattices, such as associating spinless bosons 
of the continuum with the link fields 
 on the lattice, and associating double-valued spinors  of the continuum 
with single-valued representations of the lattice point group symmetry.
To the reader  acquainted with the  so called ``topological'' twisting 
this  should all sound natural and be thought as a lattice version of it. 
And this is indeed true.  Most of this paper is a study of representation
theory of continuous and finite symmetry groups to convey this picture. 
Making the orbifold lattice-twisted continuum theory correspondence  
clear also fulfills  some curiosities on the relation between the   
two  recent independent 
approaches   on supersymmetric lattices, and in fact it reconciles them. 
This will be made more precise.

The formulation of the  twisted theories  was initiated by Witten  
in his  classic work on Donaldson theory of 
four manifolds \cite{Witten:1988ze}. Witten constructed a twisted version of 
the asymptotically free $\CN=2$ supersymmetric
Yang-Mills theory and  calculated certain topological correlators, 
both  in the ultraviolet 
 by taking advantage of  the weak coupling limit  \cite{Witten:1988ze},   
and  in  the infrared, long distance  point of view  \cite{Witten:1994cg}. 
As these  correlators are metric independent, they naturally come out to be 
the same.     
The technique for constructing twisted versions of  other extended 
supersymmetric gauge theories, such as $\CN=4$ SYM,  has been investigated 
in depth \cite{
  Yamron:1988qc, Vafa:1994tf,  Dijkgraaf:1996tz}
and the ones which are 
relevant to the discussions of orbifold lattices are due to  
Marcus \cite{Marcus:1995mq}, and   Blau and Thompson\cite{Blau:1996bx}. 

There are  two recent independent 
approaches for the construction of a nonperturbative 
regularization of the   supersymmetric gauge theories,
 and as stated earlier, one of
the primary goal of this paper is to make the relation between the two
precise. 
The first approach, the orbifold lattice,   is based on an
orbifold projection of a supersymmetric matrix model \cite{Kaplan:2005ta,
  Cohen:2003aa, Kaplan:2002wv, Kaplan:2003uh,  Cohen:2003qw }. 
The projection 
generates a lattice theory while preserving a subset of the supersymmetries 
of the target theory and benefits from the 
 deconstruction limit \cite{ArkaniHamed:2001ca}.
The  other approach,  pioneered  by Catterall,
\cite{Catterall:2001fr, Catterall:2003wd, Catterall:2004np, Catterall:2005fd,
Catterall:2006jw} uses twists of Witten type along with  
Dirac-K{\"a}hler fermions.  The  main  idea is to 
express the continuum action in a twisted  form and discretize the  
theory by keeping a subset of the nilpotent  (up to gauge transformations)
supersymmetries exact even at finite lattice spacing. The Dirac-K{\"a}hler 
fermions  have a geometric realization on the lattice and are usually
associated with $p$-cells.  
Sugino pursued an approach based on  ``balanced topological field theory
form'' \cite{Dijkgraaf:1996tz} and chose to put fermions on the sites 
\cite{Sugino:2003yb, Sugino:2004qd, Sugino:2004uv, Sugino:2006uf}.  
There are also claims that the full  twisted superalgebra can be incorporated 
to the lattice with a modified definition of Leibniz rule
\cite{D'Adda:2005zk, D'Adda:2004ia, D'Adda:2004jb}.  
 The outcomes of these two approaches 
%(based on orbifolds projection and twisted supersymmetry),   
are not identical. The reader
 may wonder why this is so, considering that  
 the orbifold lattices  produce twisted theories in the continuum.  
In a nutshell,  the difference between the two approaches can be  traced 
to the non-uniqueness of the embedding  of the  scalar  supersymmetries 
on the lattice.  As we will see, in the twisted formulation of these
theories  there is usually  more than one scalar supercharge and any linear
combination can be used on the lattice. (Also see the references 
\cite{Giedt:2006pd, Giedt:2005ae, Giedt:2004qs, 
Giedt:2003vy, Onogi:2005cz, Nishimura:2003tf, Unsal:2004cf} for 
related works.) 

One of the main outcomes of our analysis is that the discrete point 
group symmetry of the    
orbifold lattices is not a subgroup of the Lorentz group {\it per se},
 but the twisted  version of it. 
In this viewpoint,  the scalars of the physical theory turns 
out to be vectors under the twisted Lorentz symmetry, which explain 
 their appearance as the link fields.  Moreover, the spinors (double valued 
representations) of the physical 
theory transform in the single-valued integer spin representations 
of the twisted theory.  Therefore, spinors of the continuum theory are
associated with single valued lattice representations. Hence, their
appearances on the $p$-cells, sites, links, faces etc. of the hypercubic 
lattice can  be naturally  understood.  
We reach the same conclusions in two different ways. 
In the bulk of the paper, we construct twistings in terms of continuous 
groups, then translate the outcome to the lattice. 
 In a short appendix, we sketch a complementary approach. We consider 
double-valued finite groups, which are indeed the discrete spacetime and 
discrete R-symmetries. Then, we show how twisting glues objects of 
half-integer spin into integer spin  multiplets of the 
diagonal subgroup of discrete $R$ and spacetime  symmetry. 

We also explain the relation between the $A_d^*$ lattices and the twisted 
versions of $\CQ=16$ supercharge  target theories  in $d$ dimensions. 
 The $A_d^*$  lattices are the most symmetric lattices,  in particular they are
more symmetric than the hypercubic ones. This is important when considering
the quantum continuum limit and renormalization of these theories. 
The greater the symmetry of the spacetime lattice, the fewer relevant and
marginal operators will exist.
Therefore, the most symmetrical arrangements of the lattices are preferred 
to minimize the fine tunings in attaining the continuum limit.  
 The point group symmetry of $A_d^*$ lattices involves at least the
 permutation group $S_{d+1}$ (not $S_d$  as in the case of $d$ dimensional  
cubic lattice). We should emphasize that the 
group $S_{d+1}$ does not have double valued (spinor) representations at all,
even though all the target theories possess spinor representations. 
We will observe that  there is  
a  close  relation between the finite group $S_{d+1}$ and continuum 
twisted Lorentz  group and their representations.  This will be discussed 
in depth in section \ref{sec:A4star} and is one of the main results of this 
paper.

The twisted theories  emerging  from the orbifold lattices  are 
examined in the context of the topological twisting of the extended 
supersymmetric field theories.  
In four dimensions,   
the twist of $\CN=4$ is introduced   by Marcus \cite{Marcus:1995mq}. 
The three dimensional $\CN=4$ and  $\CN=8$  and two dimensional  $\CN=(8,8)$, 
$\CN=(4,4)$  theories are presented  by Blau and  Thompson
 \cite{Blau:1996bx} and are examined in more detail  in \cite{Geyer:2001yc, 
Geyer:2001qy}.  The twist of the two dimensional 
$\CN=(2,2)$  theory seems to be a new example of 
\cite{Marcus:1995mq, Blau:1996bx} type and is examined in more detail here. 
 Conversely, starting with  the
continuum  form of the twisted theory, it is possible to reverse engineer 
the hypercubic orbifold lattice by using a simple recipe given by Catterall 
\cite{Catterall:2004np}. 

\section{Maximal twisting and orbifold projection} 

In this section, we briefly review  the  twistings of  extended
supersymmetric gauge theories in the continuum  formulation on 
${\mathbb R^d}$  \cite{Witten:1988ze} and sketch its relation to orbifold 
projections of supersymmetric matrix models.
  The theories  of interest have a Euclidean rotation group $SO(d)_E$ and   
possess a   
global $R$-symmetry group $G_R$. For   six of the  theories shown in 
Table.\ref{tab:tab1},  the $R$-symmetry group possess a $SO(d)_R$ subgroup.   
Hence, the full global  symmetry of the supersymmetric  theory has a 
subgroup   
$SO(d)_E \times SO(d)_R \subset SO(d)_E \times G_R$.  
To construct the twisted theory, we embed a new  rotation group  $SO(d)'$
into the  diagonal sum of  $SO(d)_E \times SO(d)_R $,   
and declare   this  $SO(d)'$  as the new  Lorentz 
symmetry of the  theory.  \footnote{ We will not distinguish spin groups 
$Spin(n)$ from   $SO(n)$ unless otherwise specified.}
 
%%%%%%%%%%%%%%%%%%%%%%%%%%%%%%%%%%%
%%%%%%%%%%%%%%%%%%%%%%%%%%%%%%%%%%%
\setlength{\extrarowheight}{3pt}
\begin{table}[t]
\centerline{
\begin{tabular}
{|c|c|c|c|c|} \hline
Theory   & Lorentz & $\CQ=4$  &  $\CQ=8$  &  $\CQ=16$ \\ \hline
$d=2$  & $SO(2)$ &$SO(2) \times U(1)$  & $SO(4) \times SU(2)$ &  $SO(8)$ 
\\ \hline
$d=3$   &$SO(3)$ &$U(1)$  & $SO(3) \times SU(2)$    &  $SO(7)$     \\ \hline
$d=4$ &$SO(4)$  &$U(1)$  &   $SO(2) \times  SU(2)$  &   $SO(6)$   \\ \hline
\end{tabular} }
\caption{\sl  The $R$-symmetry groups of various  supersymmetric gauge 
theories obtained by  dimensionally reducing 
minimal $\CN=1$ theories from $d=4,6, 10$  dimensions. These $R$-symmetries
are the product of   the global symmetry 
due to reduced dimensions   and   the $R$-symmetry of 
the theory prior to  reduction.   
\label{tab:tab1}}
\end{table}
%%%%%%%%%%%%%%%%%%%%%%%%%%%%%%%%%%%
%%%%%%%%%%%%%%%%%%%%%%%%%%%%%%%%%%%

Since the details of each such construction are slightly different, 
let us restrict to generalities first. 
Let us assume that a fermionic field which is a spacetime spinor, 
 is  in spinor representation of $R$-symmetry group $SO(d)_R$ as well.   
Since the product of two half-integer spin is always an integer 
spin, {\it all} Grassmann odd degrees of freedom are in
integer spin representations of $SO(d)'$. We can express the fermions as a 
direct sum of scalars, vectors, i.e as  $p$-form tensors. Let us label a 
$p$-form fermion as $\psi^{(p)}$. 
In  all of our applications, the $\CQ$ many fermions of a target field theory
in $d$ dimensions    are distributed to multiplets of
$SO(d)'$ as 
\beq 
{\rm fermions} \rightarrow  \;  \frac{\CQ}{2^d} 
( \psi^{(0)} \oplus\psi^{(1)}  \oplus  \ldots  \psi^{(d)}) 
\eqn{DK}
\eeq
where the multiplicative factor up front is one, two or four.  For a given 
$p$-form, there are $  \frac{\CQ}{2^d} \binom{d}{ p}$ fermions. 
Summing over all $p$,  we obtain 
the total number of fermions in the target theory:
%\beq
$\frac{\CQ}{2^d} \sum_{p=0}^{d}  \binom{d}{p}= \CQ 
%\eeq
$

Turning to  Grassmann even fields, 
the gauge bosons  $V_{\mu}$ transforming as $(d, 1)$   and  the spacetime 
scalars   $S_{\mu}$
transforming as $(1,d)$  under the   $SO(d)_E \times SO(d)_R $ level. 
Both transform as vectors $(d)$ under the    $SO(d)'$.  If there are more then 
$d$ scalars in the  untwisted theory, they become either $0$-forms or 
$d$-forms under $SO(d)'$.

This type of twist is sometimes referred as {\it maximal twist} as it
involves the twisting of the full Lorentz symmetry group as opposed 
to twisting its subgroup.  In this sense,  
the  four dimensional  $\CN=2$ theory  can only admit a {\it half twisting}
 as its $R$-symmetry 
group is not as large as $SO(4)_E$ \cite{Witten:1988ze}. The other two
theories, $\CN=1$ in $d=4$ and  $\CN=1$ in  $d=3$ shown in 
Table.\ref{tab:tab1} do not admit a nontrivial twisting 
as there is no nontrivial homomorphism from their Euclidean rotation group
to their $R$-symmetry group. 

The action  expressed in terms of  the representation  of the  
twisted Lorentz  group $SO(d)'$   instead of the ones of the usual 
Lorentz symmetry, is called twisted action.   The twisted version  
can be expressed as a sum of $Q$-exact and $Q$-closed terms, 
where $Q$-is  the supersymmetry associated with scalar supersymmetry 
transformation.
\footnote{One can make 
this theory topological by interpreting the scalar supercharge $Q$ as a BRST
operator \cite{Witten:1988ze}. Even without doing so,  
one can still say that the physical theory has a set of 
topological observables, appropriately defined correlators of the 
twisted operators.}  
As it is well know, so long as the usual  Lorentz symmetry is not gauged,
i.e., on flat spacetime, the twisted  theory is merely a 
rewriting  of the physical theory, and 
indeed possess all the supersymmetries of the physical theory.  
\footnote{
In fact, if the base space of the theory is an arbitrary 
 $d$-dimensional curved  manifold $M^{d}$, then 
only the scalar  supercharge is preserved. It is somehow peculiar
that the discretized background spacetime (lattice)  also  respects 
the one and same nilpotent  scalar supersymmetry.}   

The main point of this twist is that none of the  degrees of freedom are 
spinors under $SO(d)'$.  Both   bosons and 
fermions are in  integer spin representations. They are $p$-form tensors 
 of $SO(d)'$. 
This particular form of the twisted theory is the bridge to orbifold
lattices. Given such a twisted theory,  
it is natural to associate a $p$-form continuum
field  with a $p$-cell field on the hypercubic lattice. 
This is exactly what an orbifold lattice does. The orbifold projection 
places the fermions to sites, links, faces, i.e, to $p$-cells. 
This is in 
agreement  with our expectation from the twisted rotation symmetry
$SO(d)'$.  
On the orbifold lattice,  there are also complex bosons (complexification 
of $S_{\mu}$ and $V_{\mu}$ as   $(S_{\mu} \pm iV_{\mu})/\sqrt 2$)  
associated with  oppositely oriented links  and  
certain fields associated with $p$-cells.  
We refer the reader to ref.\cite{Kaplan:2002wv, Kaplan:2005ta} for a 
detailed explanation of the orbifold projection and ${\bf r}$-charge 
assignments. 
By using the analysis of ref.\cite{Kaplan:2002wv}, we see that 
${\bf r}$-charge assignment    is intimately related to 
how a field transforms in the continuum. Mainly, 
the total number of 
nonzero components of the ${\bf r}$-charge is the degree $p$ of the tensor 
representation of $SO(d)'$. The    signs of components of ${\bf r}$ 
determine the orientation  of the corresponding lattice field. 
For example, on a $d=2$ dimensional square lattice, we
associate fermions with 
${\bf r}=(0,0)$ with  0-cell, ${\bf r}=( 1,0)$ with 
1-cells in $\bfe_1$ direction,    ${\bf r}=( 0, 1)$ with 1-cells in  
 $\bfe_2$-direction and ${\bf r}=( -1, -1)$ with a 2-cell field in 
 $-\bfe_1 -\bfe_2$ direction. These respectively become zero, one and two
 form tensor fermions as in \Eq{DK} under the continuum $SO(d)'$.

One may ask how does  these orbifold projections
know about the representations of the twisted group. 
Recall that 
the  ${\bf r}$-charges are given in  an appropriate abelian 
subgroup of the {\it full}  $R$-symmetry group of a zero dimensional 
 matrix model.  
This matrix  model  is  obtained by dimensionally 
reducing the target theory to zero dimension  and possesses {\it at least}  
an $SO(d)_E \times  G_R$ $R$-symmetry group. 
The  $G_R$ is the $R$-symmetry prior to reduction and  more importantly,
$SO(d)_E$, which used to be  the Lorentz symmetry of the target theory,  is an 
$R$-symmetry of matrix theory.   The full $R$-symmetry group 
of the matrix theory  is in general  larger than  $SO(d)_E \times  G_R$.
For example, for $\CN=4$ SYM theory reduced from $d=4$ to $d=0$ dimensions 
has a manifest    $SO(4)_E \times  SO(6)_R$ $R$-symmetry, but it clearly  
enhances  to $SO(10)_R$.    
The choice of  ${\bf r}$-charges mixes the global Lorentz $R$-symmetry 
$SO(d)_E$   with  $G_R$ in a profound way.  
It is in fact a form of twisting.  Let us again
restrict to $\CN=4$ SYM theory and its reduced version. The reduced version 
has an $SO(10)_R$. For hypercubic lattices, the $SO(10)_R \rightarrow SO(8)
\times SO(2) \rightarrow SU(4) \times U(1) \times U(1) $  branching  plays a
fundamental role. In fact, the $\bfr$  charges is embedded in  $U(1)^4$
subgroup of $SU(4) \times U(1) \times U(1)$ and the intact $U(1)$ remains to
be an $R$-symmetry of lattice theory.  One may wonder what does this 
 $SU(4)$ has anything to do with the diagonal $SO(4)'$ of continuum theory.  
The answer is  somewhat subtle.  
The  lattice  which is obtained by the orbifold 
projection has a  finite nonabelian  point group symmetry, which is 
the Weyl group of $SU(4)$, i.e.,  
$Weyl(SU(4))=S_4$ isomorphic to the permutation  group 
$S_4$. And in fact, this Weyl group is the discrete 
subgroup of diagonal $SO(4)'$, i.e, $S_4 \subset SO(4)'$.  
Under a conveniently chosen 
abelian subgroup of full $R$-symmetry,  the fermions carry integer 
charges as bosons and they form supermultiplets. These multiplets transform in 
 representation of non-abelian point group symmetry (or equivalently 
Weyl group). 
 
The proper understanding of the more symmetric $A_d^*$ lattices, which arise 
for $\CQ=16$ supercharge target theories in $d$ dimensions,  
from the twisted supersymmetry  viewpoint is a little bit more involved, but
is a  worthy endeavor.  The $A_d^*$  lattices are the most symmetric
lattices,  
and the
greater the symmetry of the spacetime lattice, the fewer relevant and
marginal operator will exist. 
 The point group symmetry of $A_d^*$ involves 
 the Weyl group of $SU(d+1)$, rather than $SU(d)$. For example, the highly
 symmetric lattice $A_4^*$ for $\CN=4$ SYM theory has an $S_5= Weyl(SU(5))$
 point group  symmetry, which is much larger than the point group symmetry 
of hypercubic lattice.  
The   classification of  the fields 
on the $A_4^*$ lattice under the point group symmetry  
is discussed in detail in the next sections. 
 As we will see, there is  
a  close  relation between $Weyl(SU(d+1))=S_{d+1}$ and continuum 
twisted rotation  group $SO(d)'$ and their representations.  

This line of reasoning teaches us that  
the point group symmetry of the lattice is {\it not} a subgroup of the 
Euclidean Lorentz  group, but in fact a discrete subgroup of the twisted 
rotation group $SO(d)'$. 
 In the continuum, the orbifold lattice 
theory becomes 
the twisted version of the desired target field theory.
The change of variables which takes the twisted form to the canonical 
form  essentially  undoes the twist.
                                                    
\section{Marcus's twist   of $\CN=4$ SYM in $d=4$}
\label{sec:marcus}
There are various possible  twists of the $\CN=4$ SYM theory in 
four dimensions \cite{Yamron:1988qc,Vafa:1994tf, Marcus:1995mq }. 
The one we will consider and which emerges out of the orbifold lattice 
naturally is due to Marcus. Here, we briefly outline the twisting 
procedure. One  interesting property of this 
twisting is that it admits a superfield formulation.

The $\CN=4$ SYM theory in $d=4$ dimensions possesses  a global Euclidean 
Lorentz symmetry 
$SO(4)_E \sim   SU(2) \times SU(2)$, a global $R$-symmetry group  
$SO(6) \sim SU(4)$. The $R$-symmetry contains  a subgroup $SO(4)_R \times U(1)$.
To construct the twisted   theory, we take  the diagonal
sum of $SO(4)_E \times SO(4)_R $ and declare it  the new rotation
group. Since the $U(1)$ part of the symmetry group is undisturbed, 
it remains as  a  global $R$-symmetry of the twisted theory.    

Under the   
 $ G= \big( SU(2) \times SU(2) \big)_E \times 
\big( SU(2) \times SU(2) \big)_R $  symmetry, the fermions transform as 
$(2,1,2,1) \oplus  (2,1,1,2) \oplus 
  (1,2,1,2) \oplus    (1,2,2,1) $.
These fields, under $G'=SU(2)' \times SU(2)' \times U(1)$ (or  under 
$SO(4)' \times U(1)$) transform 
as\footnote{Twice of the $U(1)$  charge is usually called the ghost number in 
the topological counterpart of this  theory.}    
\beq 
{\rm fermions} \; && \rightarrow (1,1)_{\frac{1}{2}} \oplus 
(2,2)_{-\frac{1}{2}} 
\oplus [(3,1) \oplus (1,3)]_ {\frac{1}{2}}
\oplus  (2,2)_{-\frac{1}{2}}  \oplus (1,1)_{\frac{1}{2}}  \cr
&&  \rightarrow 1_{\frac{1}{2}} \oplus 
4_{-\frac{1}{2}} 
\oplus 6_ {\frac{1}{2}}
\oplus  4_{-\frac{1}{2}}  \oplus 1_{\frac{1}{2}} \; .
\eeq
 The magic of this particular embedding 
is clear. There are two spin zero fermions, and  all the fermions are now 
in the integer spin representation  of the 
twisted Lorentz symmetry  $SO(4)'$.  
They transform as  scalars, vectors, and 
higher rank $p$-form tensors.   We parametrize these 
Grassmann valued tensors, accordingly, 
 $( \lambda,  \psi^{\mu},  \xi_{\mu \nu}, 
\xi^{\mu \nu \rho},  \psi_{\mu \nu \rho \sigma})$. 

 The gauge boson $V_{\mu}$ which 
transform as $(2,2,1,1)$ 
under the  group $G$  becomes $(2,2)$ under $G'$. 
Similarly, four of the  scalars $S_{\mu}$ transforming as $(1,1, 2 ,2)$ 
are elevated  to the same footing as the gauge boson and transform as  
$(2,2)$  under twisted rotation group.
The  complexification of the two vector plays a more fundamental role in the
formulation. We therefore define   the complex vector fields 
\footnote{Throughout this
 paper,   $\mu, \nu, \rho, \sigma
\ldots  $   are $SO(d)'$ or $d$-dimensional 
hypercubic  indices  and summed over $1, \ldots d$. The indices $ m, n,
\ldots$ are indices for permutation group $S_{d+1}$  (for $A_d^*$ lattices)  
and   are summed over  $1,\ldots ,(d+1).$ }
\beq
z^{\mu}= (S^{\mu} + i V^{\mu})/ \sqrt 2,  \qquad 
 \mybar z_{\mu}= (S_{\mu} - i V_{\mu})/ \sqrt 2  \qquad \mu=1, \dots, 4
\eqn{vector}
\eeq 
Since there are two types of vector  fields, there are indeed two types of 
complexified gauge covariant derivative appearing in the formulation. 
These are holomorhic and antiholomorphic covariant  derivatives   
\beq
%D_{\mu}\, \cdot  = \partial_{\mu} \cdot  + i [ V_{\mu}, \, \cdot \,], 
%\qquad 
 \CD^{\mu}\, \cdot  = \partial^{\mu} \cdot + \sqrt 2 [ z^{\mu}, 
  % S_{\mu} +  i  V_{\mu},
  \, \cdot \,
], 
\qquad 
 {\mybar \CD}_{\mu} \, \cdot  = -\partial_{\mu} \cdot  +  \sqrt 2 
[ \mybar z_{\mu}
%S_{\mu} -i V_{\mu}
 , \, \cdot \, ] \, , 
\eqn{cov}
\eeq
Only three combination of the covariant derivatives (similar to the  $F$-term
and  $D$-term in  the $\CN=1$ gauge theories)  appear in the
formulation. These are 
\beq
\CF^{\mu \nu}&&= -i [\CD^\mu, \CD^\nu] 
%=  \sqrt2 ( \partial^{\mu} z^{\nu} -  \partial^{\nu} 
%z^{\mu} + i \sqrt 2 [ z^{\mu},  z^{\nu}] )
= F_{\mu\nu} -i [S_{\mu}, S_{\nu}] 
-i (D_{\mu}S_{\nu}- D_{\nu}S_{\mu}) \cr  
 {\mybar \CF}_{\mu \nu}&&= 
-i [{\mybar \CD}_\mu, {\mybar \CD}_\nu] =  F_{\mu\nu} -i [S_{\mu}, S_{\nu}] 
+i (D_{\mu}S_{\nu}- D_{\nu}S_{\mu}) \cr
(-id) &&= \half  [\mybar {\cal D}_{\mu}, {\cal D}^{\mu}] + \cdots   = 
-D_{\mu}S_{\mu} + \cdots
\eqn{strength}
\eeq
where $D_{\mu}\, \cdot  = \partial_{\mu} \cdot  + i [ V_{\mu}, \, \cdot \,]$
is the usual covariant derivative and   $F_{\mu\nu} = -i [D_\mu, D_\nu]$
is the nonabelian  field strength. 
%$F_{\mu\nu} = -i [D_\mu, D_\nu] = \partial_{\mu} V_{\nu} -  \partial_{\nu} 
%V_{\mu} + i[ V_{\mu},  V_{\nu}]$. 
The field strength $\CF^{\mu \nu}(x) $ is holomorphic, it only
depends on complexified vector field $z^{\mu}$   and not on $\mybar
z_{\mu}$. Likewise,   $\mybar \CF_{\mu \nu}$  is anti-holomorphic. 
The $(-id)$ will come out of the solutions of equations of motion for
auxiliary field $d$ and 
dots stands for possible scalar  contributions. 
These combination arises from all of the  orbifold lattice constructions,  
and is one of the reasons for considering this type of twist. 

Finally, the two other scalars remains as scalars  under the twisted 
rotation  group. Since one of the scalars is the  superpartner 
(as will be seen below)  
of  the four form fermion,  we label them as  
$(z_{\mu \nu \rho \sigma}, \mybar z^{\mu \nu \rho \sigma} )$. 
To summarize,  the bosons transform under $G'$ as   
\beq 
{\rm bosons} \rightarrow  
z_{\mu \nu \rho \sigma} \oplus  z^{\mu} \oplus  \mybar z_{\mu} \oplus 
\mybar z^{\mu \nu \rho \sigma}   \; \rightarrow  
[ (1,1)_1 \oplus (2,2)_0 +  (2,2)_0 + (1,1)_{-1} ]
\eqn{bosons}
\eeq

As can be seen easily from the decomposition of the fermions, there are two
Lorentz singlet supercharges $(1,1)$ under the twisted Lorentz group and 
either of these (or their linear  combinations) can be used to write down 
the Lagrangian  of the  four dimensional theory in ``topological'' form. 
The difference  in lattices obtained in  
\cite{Catterall:2005eh, Catterall:2005fd,  Sugino:2004uv} and 
orbifold lattices  \cite{Kaplan:2005ta} is  tightly related to the choice of
the scalar supercharge, and  this  will be further discussed in 
section \ref{sec:othertwist}. 
Here, we use the spin zero supercharge associated with $\lambda$ (motivated by
the orbifold lattice). This produces the transformations given by  
\cite{Marcus:1995mq}. 

The  continuum off-shell  supersymmetry transformations are given by  
\begin{eqnarray}
&&Q \lambda =  -id,   \qquad    Q d = 0 \cr 
&& Q z^{\mu} =   \sqrt 2 \, \, \psi^{\mu}, \qquad  Q \psi^{\mu}=0 \cr
&& Q \mybar z_{\mu} = 0  \cr
&&Q \xi_{\mu\nu} = -i \mybar \CF_{\mu \nu} \cr
&&Q \xi^{\nu \rho \sigma} = \sqrt 2 \, {\mybar \CD}_{\mu}  
\mybar z^{\mu \nu \rho \sigma}  \cr
&&Q z_{\mu \nu \rho \sigma} = \sqrt{2} \psi_{\mu \nu \rho \sigma}, \qquad 
Q \psi_{\mu \nu \rho \sigma} = 0    \cr 
&&Q \mybar z^{\mu \nu \rho \sigma} =0  
\eqn{QAoffshell}
\end{eqnarray}
where $d$ is an auxiliary field introduced for the  off-shell completion of 
the  supersymmetry algebra.  
Clearly, the scalar    supercharge is nilpotent  
\beq Q^2 \; \cdot  = 0. \eeq
owing to the anti-holomorphy of $\mybar \CF_{\mu \nu}$ etc.    
The fact that the subalgebra  ($Q^2=0$) does not 
produce any spacetime translations makes it possible to carry it easily onto
the lattice.
The exact nilpotency, as opposed to being nilpotent modulo gauge
transformation  has a technical advantage.  It admits a rather 
exotic superfield formulation of the target supersymmetric field theory which 
will be discussed in the next section. 

The twisted  Lagrangian  may be written as a sum of $Q$-exact and  $Q$-closed 
terms:
\beq
g^2 \CL = && \CL_{exact} +\CL_{closed} =  \CL_1 +  \CL_2 +   \CL_3 =  
Q {\widetilde \CL_{exact}} +  \CL_{closed},
\eqn{LT41} 
\eeq 
where $g$ is coupling constant and 
 $\widetilde \CL_{exact}= {\widetilde \CL}_{e,1} + 
{\widetilde \CL}_{e,2} $ is given by  
\beq
&& {\widetilde \CL}_{e,1} =    \Tr  \Big( \lambda ( \half id + \half 
[ \mybar {\cal D}_{\mu}, {\cal D}^{\mu} ] +
\textstyle{\frac{1}{24}}[\mybar z^{\mu\nu \rho \sigma}
, z_{\mu\nu \rho \sigma}] ) \Big) \cr
&& {\widetilde \CL}_{e,2}= 
  \Tr  \Big( 
\fourthi \xi_{\mu\nu} \CF^{\mu\nu} +  \textstyle{\frac{1}{12 \sqrt2}}
\xi^{\nu \rho \sigma} {\cal D}^{\mu} z_{\mu \nu \rho \sigma} 
\Big) 
\eeq
and   $\CL_{closed}$ is given by 
\beq
 \CL_{closed}= \CL_3 = \Tr 
\half \xi_{\mu\nu }  \mybar {\cal D}_{\rho} \xi^{\mu \nu \rho} +  
\textstyle{\frac{\sqrt2}{8}} \;  \xi_{\mu\nu } [\mybar 
z^{\mu \nu \rho \sigma}, \xi_{\rho \sigma}] \qquad
\eeq
By using the transformation properties  of fields and 
the equation of motion auxiliary field $d$  
\beq
(-id)= \half  [ \mybar {\cal D}_{\mu}, 
{\cal D}^{\mu} ] +
\textstyle{\frac{1}{24}}[\mybar z^{\mu\nu \rho \sigma}
, z_{\mu\nu \rho \sigma}] \; ,
\eeq
we obtain the Lagrangian expressed in terms of propagating degrees of 
freedom: \footnote{Notice that the splitting  of the exact terms in 
Lagrangian  into $\CL_1$ and $\CL_2$  is not identical to the one used 
by Marcus.  The reason for the above splitting lies in the symmetries of the
cut-off theory ($A_d^*$ lattice theory)  that will be discussed later.}
\beq 
 &&  \CL_1 =   \Tr  \Big( \half (  \half [\mybar {\cal D}_{\mu} , 
 {\cal D}^{\mu}] + \textstyle{\frac{1}{24}}[\mybar z^{\mu\nu \rho \sigma}
, z_{\mu\nu \rho \sigma}])^2 +   
\lambda ( \mybar {\cal D}_{\mu} \psi^{\mu} +   
\textstyle{\frac{1}{24}}[\mybar z^{\mu\nu \rho \sigma}
, \psi_{\mu\nu \rho \sigma}]) \Big)  \cr 
&&  \CL_2 =   \Tr  \Big(  \fourth \mybar {\CF}_{\mu\nu} {\CF}^{\mu\nu}    
+   \xi_{\mu\nu}  {\cal D}^{\mu} \psi^{\nu} 
+ {\textstyle \frac{1}{12}} |{\cal D}^{\mu} z_{\mu \nu \rho \sigma}|^2
+  {\textstyle \frac{1}{12}} 
 \xi^{\nu \rho \sigma} {\cal D}^{\mu} \psi_{\mu \nu \rho \sigma} 
+  {\textstyle \frac{1}{6 \sqrt 2 }}
 \xi^{\nu \rho \sigma} [\psi^{\mu}, z_{\mu \nu \rho \sigma}]
\Big)   \cr 
&&  \CL_3 = \Tr \Big(
\half \xi_{\mu\nu }  \mybar {\cal D}_{\rho} \xi^{\mu \nu \rho} +  
\textstyle{\frac{\sqrt2}{8}} \;  \xi_{\mu\nu } [\mybar 
z^{\mu \nu \rho \sigma}, \xi_{\rho \sigma}] 
\Big) \; .
\eqn{LT4}
\eeq
The $Q$-invariance of the  $\CL_{exact}$ is obvious and  
 follows from supersymmetry algebra $Q^2=0$.
To show the invariance of $Q$-closed term  
requires the  use of the Bianchi 
(or Jacobi identity for covariant derivatives)  identity   
\beq
\epsilon^{\sigma \mu\nu \rho } \mybar {\cal D}_{\mu}  \mybar \CF_{\nu \rho}=
\epsilon^{\sigma \mu\nu \rho } [\mybar {\cal D}_{\mu}, [ 
\mybar {\cal D}_{\nu},  \mybar {\cal D}_{\rho} ]] = 0
\eqn{Jacobi}
\eeq
and similar identity involving scalars. 
The action is expressed in terms of the  twisted Lorentz multiplets, and  
the   $SO(4)'\times  U(1)$ symmetry is manifest. 
The Lagrangian  \Eq{LT4} emerges from the hypercubic and 
$A_4^*$ lattice action 
at the tree level.
This will be discussed after the following digression 
to superfield formulation of Marcus's twist. 

\subsection{The $\CQ=1$ (twisted) superfields formulation of $\CN=4$ SYM}
In this section, we introduce a superfield notation  for the 
the  twisted  $\CN=4$ SYM theory.  The superfields 
are $SO(4)'$ multiplets.   
This is so since the manifest supersymmetry  is a scalar and   exactly 
nilpotent. Consequently, 
different components of a multiplet (unlike the usual supersymmetry 
multiplets in four dimensions) reside in the same representation of 
twisted rotation group.  

The supermultiplets are all in   integer spin representations  of $SO(4)'$. 
The 
superfields  are  a scalar fermi multiplet 
${\bf \Lambda}(x)$  transforming as $(1)_{\frac{1}{2}}$,    
a vector multiplet  ${\bf Z}^{\mu}(x)$  transforming as $(4)_0$, 
a two-form  
fermi  multiplet
 ${\bf \Xi}_{\mu \nu}(x)$ transforming as  
 $(6)_{\frac{1}{2}}$, 
a  three-form  fermi multiplet  ${\bf \Xi}^{\mu \nu \rho}(x)$ in  
$(4)_{-\frac{1}{2}}$, and a four form ${\bf Z}_{\mu \nu \rho \sigma } (x)$ 
in  $(1)_1$ representation. 
 There are also two types of supersymmetry singlets,  
a vector ${\mybar  z}_{\mu}(x)$  in   $(4)_0$ 
and a four form  ${\mybar  z}^{\mu \nu \rho \sigma}(x)$ transforming as 
 $(1)_{-1}$. 
The  scalar  $\CQ=1$ off-shell
supersymmetry transformations  can then be realized in terms of these  
superfields as 
\beq
{\bf \Lambda}(x) &=& \lambda (x)  -\theta   id (x), \cr
{\bf Z}^{\mu}(x)  &=&  z^{\mu}(x) + \sqrt 2  \theta \psi^{\mu} (x), 
\qquad \mybar z_{\mu} (x) , \cr
{\bf \Xi}_{\mu \nu}(x) &=& \xi_{\mu \nu}(x) -i\theta  {\mybar \CF}_{\mu \nu}(x), \cr
{\bf \Xi}^{\nu \rho \sigma}(x) &=& 
\xi^{\nu \rho \sigma}(x) + {\sqrt 2} \theta {\mybar \CD}_{\mu}\mybar 
z^{\mu \nu \rho \sigma} (x),  \cr
{\bf Z}_{\mu \nu \rho \sigma } (x)  &=&  z_{\mu \nu \rho \sigma}(x) + 
\sqrt 2  \theta \psi_{\mu \nu \rho \sigma} (x),  
 \;\;  \qquad   
 \;\; {\mybar z}^{ \mu \nu \rho \sigma} (x). 
\eeq
These superfields should be useful in formulating the $\CN=4$ SYM 
 not only on $\mathbb R^4$, but on arbitrary  curved four-manifold
 $M^4$, mainly because they are based on scalar supersymmetry. 
 By introducing the  super-covariant derivative;   
\beq
{\pmb {\cal D}}^{\mu}= \partial^{\mu} + \sqrt 2 {\bf Z}^{\mu} =  \CD^{\mu} +  
2 \theta {\psi}^{\mu} 
\eeq
we can  also  define the field strength multiplet  
\beq
{\pmb {\CF}  }^{\mu \nu} = -i [ {\pmb {\CD}}^{\mu}, {\pmb {\CD}}^{\nu}] = 
{\CF}^{\mu \nu}-2i \theta (\CD^{\mu} \psi^{\nu} -  \CD^{\nu} \psi^{\mu})
\eeq
transforming as  $(6)_{0}$ under $SO(4)' \times U(1)$. 
In terms of the $\CQ=1$  superfields, 
the action of the $\CN=4$ SYM  theory on $\mathbb R^4$ can be expressed as 
\beq
S = \frac{1}{g^2} \Tr  \int d^4 x \;  d \theta 
\bigg(&& -\frac{1}{2} 
{\bf \Lambda} 
{\partial}
_{\theta} {\bf \Lambda}  - {\bf \Lambda} (  \frac{1}{2} [\mybar \CD_{\mu}, 
{\pmb \CD}^{\mu}] 
+  \frac{1}{24} \; [\mybar z^{\mu \nu \rho \sigma}, {\bf Z}_{\mu \nu \rho \sigma}]) \bigg. \cr &&
\bigg.  + 
 \frac{i}{4}
{\bf \Xi}_{\mu \nu} {\pmb \CF}^{\mu\nu} 
% \bigg. \cr && \bigg. 
+  \frac{1}{12\sqrt2}\;
{\bf \Xi}^{\nu \rho \sigma }  
{\pmb \CD}^{\mu} {\bf Z}_{\mu \nu\rho \sigma}   \bigg) 
\cr&&
+  \frac{1 }{2 }    
 {\bf \Xi}_{\mu \nu} 
 \;  \mybar \CD_{\rho} \;  
  {\bf \Xi}^{ \mu \nu\rho }   
 +
\frac{\sqrt 2}{8}   {\bf \Xi}_{\mu \nu}
[\mybar z^{\mu \nu\rho \sigma} ,  {\bf \Xi}_{\rho \sigma} ] 
 \qquad 
\eqn{LS4}
\eeq
The last line is not integrated  over the superspace  and is the $Q$-closed
term discussed above. Its  $\theta$ component vanishes because of Jacobi
identities, hence it is supersymmetric.  Notice that the three lines of this 
action respectively corresponds to $\CL_1,  \CL_2, \CL_3$ in  \Eq{LT4}.

\subsection{Hypercubic  lattice}
The action \Eq{LT4}, or equivalently \Eq{LS4}, expressed in terms of 
integer spin representations ($p$-forms) of  $SO(4)'$ arises 
naturally from orbifold  lattices \cite{Kaplan:2005ta, Unsal:2005us}. 
Recall that  the  fundamental cell of the hypercubic lattice 
contains one site, four links, six faces, 
four cubes and one hypercube, collectively named as $p$-cells.
A  $p$-form tensor fermion is associated with a $p$-cell 
on the hypercubic  lattice. 
The complex vectors of $SO(4)'$ are 
associated  with the link fields. Finally, the two scalars  (four-forms) 
are associated with 
the   four-cell as can be deduced from the supersymmetry algebra. 

The  action \Eq{LS4} with manifest scalar supersymmetry (in fact,  possessing 
all sixteen supersymmetries) admits a discretization to a  hypercubic lattice 
in which one preserves the  scalar supercharge.  The hypercubic 
lattice action  is given in \cite{Unsal:2005us}. The rules of latticization 
are natural  and  given by Catterall (except the rule which requires
complexification of the fields. Our bosons and fermions are already 
complex and oriented.) 
\cite{Catterall:2004np, Catterall:2005eh}.  For our purpose, 
it suffices to understand the transformations given in \Eq{QAoffshell}.  
The local transformations in  \Eq{QAoffshell} remain the  same, 
modulo the trivial  substitution of spacetime position $x$ with 
a discrete lattice position  
index  ${\bf n}$.  There are two types of
semi-local transformation.  The first one is 
$Q \xi_{\mu\nu}(x) = -i \mybar \CF_{\mu \nu}(x) $. This translates  to 
$Q \xi_{\mu\nu, \bfn} = -2 (\mybar z_{\mu, \bfn +  \bfe_\nu}  
\mybar z_{\nu, \bfn} - \mybar z_{\nu, \bfn +  \bfe_\mu}  
\mybar z_{\mu, \bfn})  $. The right hand side is the square root of the usual 
Wilson plaquette term. Similarly,  $\CF^{\mu \nu}(x)$ becomes  $z^{\mu}_{\bfn}  
 z^{\nu}_{ \bfn + \bfe_\mu } -  z^{\nu}_{ \bfn}  
z^{\mu}_{ \bfn + \bfe_\nu }$
\footnote{Recall that the usual Wilson action may also  be written as 
$S= \sum_{\bfn}\Tr|U_{\mu, \bfn}  
 U_{\nu, \bfn + \bfe_\mu } -  U_{\nu, \bfn}  
U_{\mu, \bfn + \bfe_\nu }|^2 $  where the  quantity in  modulus
is the field strength and is indeed the  square root of a  plaquette.}.
The second transformation 
$Q \xi^{\nu \rho \sigma}(x) = \sqrt 2 \, {\mybar \CD}_{\mu}  
\mybar z^{\mu \nu \rho \sigma}(x) $ translates into  
$Q \xi^{\nu \rho \sigma}_{\bfn} =  2 
({\mybar z}_{\mu,  \bfn- \bfe_{\mu \nu \rho \sigma} }
\; \mybar z^{\mu \nu \rho \sigma}_{ \bfn} - \mybar z^{\mu \nu \rho
  \sigma}_{\bfn + \bfe_{\mu}} \; {\mybar z}_{\mu,\bfn}) 
$ where  $\bfe_{\mu \nu \rho \sigma}= \sum_{\zeta} \bfe_{\zeta}$ and 
$(\bfe_{\mu})_{\nu}= \delta_{\mu\nu}$ are the cartesian unit vectors. 
It is appropriate to parametrize the complex link fields $z^{\mu}$  as 
\beq 
z^{\mu}_{\bfn} = \frac{1}{\sqrt 2 a} e^{a(S_{\mu, \bfn} + i V_{\mu, \bfn})},
\eeq
where $a$ is the lattice spacing. Substituting these into, for example, 
 $\mybar \CF_{\mu \nu} {\CF}^{\mu\nu}$ produce a  complexified Wilson action.
This 
parametrization  differs from the ones used in  
\cite{Unsal:2005yh, Kaplan:2005ta}. However, the difference in the 
continuum is  in the  irrelevant operators, suppressed 
by powers of the lattice spacing.  This prescription generates the lattice 
actions discussed in detail in   \cite{Unsal:2005us, Kaplan:2005ta} 
and  we  will not duplicate it here.  
Instead, we want to comment on the
emergence of large global chiral symmetries,  the $R$-symmetry, 
in the continuum of orbifold lattices.

In lattice QCD,  Poincar{\'e} invariance  emerges in the continuum 
without any fine tuning, due to the point group symmetry, discrete
translation  symmetry and gauge invariance. The Poincar{\'e} violating 
relevant and marginal operators are usually forbidden due to these symmetries,
and we recover the Poincar{\'e} invariant target theory.  
In our case, the continuum limit of the hypercubic lattice 
at tree level, by construction,    
reproduces the target  theory with a  twisted Lorentz  invariance  $SO(4)'$.  
The  $U(1)_R$ symmetry is exact on  the hypercubic   lattice, and hence 
it is exact in   the  continuum.  
The $SO(4)'\times U(1)$ invariant target theory is   
a redefinition of the physical $\CN=4$ SYM
theory, which possesses  a Lorentz symmetry group and a large
$R$-symmetry, $SO(4)\times SO(6)$. The twisting obscures the large 
$R$-symmetry.  However, knowing how   $SO(4)'$ arises, we see that there is
really an $SO(4)_E \times SO(4)_R$ behind what appears to be a twisted
Lorentz symmetry. This means the large $R$-symmetry group arises from the
lattice hand in hand with Lorentz symmetry. This happens to be so since the
point symmetry group of the lattice is a subgroup of the diagonal subgroup 
of  $SO(4)_E \times SO(4)_R$. Of course, the full $R$-symmetry is $SO(6)$ and 
the above tree level argument only explains $SO(4)_R \times U(1)$ 
subgroup of it.  We will, nevertheless, be content with it.

\subsection{What does the $A_4^*$ lattice  knows  about   twisting?}
\label{sec:A4star}
In this section, we want to explain an elegant relation  between  $A_4^*$ 
lattice and the  twisted continuum theory \Eq{LT4}\footnote{This techniques 
used in this section borrows from  unpublished notes of David B. Kaplan 
on $A_3^*$  lattices for spatial lattice construction of $\CN=4$ SYM in the 
context of renormalization. I would
like to thank him for sharing them with me.}.  The $A_4^*$ lattice for 
$\CN=4$ SYM theory is introduced in ref.\cite{Kaplan:2005ta} and arises 
as the  most  symmetrical lattice arrangement in the moduli space of orbifold
lattice theory. 
In particular, it is more symmetric than the  hypercubic lattice. 
Higher symmetry 
is  an important  virtue when the  renormalization and the quantum continuum 
limit  of the lattice theory is addressed. When considering the  
radiative corrections, the  relevant and
marginal operators   will be restricted by the symmetries of the underlying
theory.  Therefore,  fewer  relevant and marginal  operators will exist 
for the  more symmetric spacetime  lattice. 
For lower dimensional examples, the
combination of lattice point group symmetry, the exact supersymmetry and
superrenormalizibility are used to show that the desired target theories 
are attained with no or  few  fine tunings at the quantum level.   
We hope that the techniques of 
this section  can eventually be used in addressing the  important problem of 
renormalization of $\CN=4$ in $d=4$ dimensions.
Our aim here is  different, and  in fact  more modest - namely showing  
the relation  between the $A_4^*$  lattice and Marcus's twist.  Our 
 analysis  is  at the tree level. We show this relation   
 by finding the irreducible  representations of
the point group symmetry of the lattice action, and  by identifying them with
the ones of  the  twisted Lorentz group  $SO(4)'$.

We have already seen the  
field distribution  on the hypercubic lattice and  identified lattice 
$p$-cell  fields with  $p$-form tensors in the continuum. The situation for 
$A_4^*$ is a little more subtle  and requires basic   representation 
and character theory  for finite groups. 
The generalization to other dimensions for which target theory is  $\CQ=16$  
and lattice is $A_d^*$ is obvious.

The $A_4^*$  lattice is generated by the fundamental weights, or 
equivalently  by the weights of defining representation  $SU(5)$. A
specific  basis  for $A_4^*$ lattice is given in the form of five, four
dimensional lattice vectors: 
\beq
  {\bf e_1} &=&  {\textstyle (  \frac{1}{\sqrt 2},  \frac{1}{\sqrt 6}, 
\frac{1}{\sqrt{12}}, \frac{1}{\sqrt{20}})}  \cr
{\bf e_2} &=&  {\textstyle (-\frac{1}{\sqrt 2},  \frac{1}{\sqrt 6},
\frac{1}{\sqrt{12}}, \frac{1}{\sqrt{20}}) } \cr
{\bf e_3} &=&   {\textstyle (0,  -\frac{2}{\sqrt 6},
\frac{1}{\sqrt{12}}, \frac{1}{\sqrt{20}})} \cr
{\bf e_4} &=& {\textstyle (0, 0,
-\frac{3}{\sqrt{12}}, \frac{1}{\sqrt{20}}) } \cr
{\bf e_5} &=&  {\textstyle (0, 0,
0, -\frac{4}{\sqrt{20}})}.  
\eqn{latvec}
\eeq 
These vectors satisfy the relations
\beq
\sum_{m=1}^5 \bfe_m = 0\ ,\qquad  \bfe_m\cdot \bfe_n =
\left(\delta_{mn}-\frac{1}{5}\right)\ ,\qquad \sum_{m=1}^5
(\bfe_m)_\mu (\bfe_m)_\nu = \delta_{\mu\nu}\ .
\eeq
The lattice vectors \Eq{latvec}  connect the center of a 4-simplex to its
five corners and  are simply related to the $SU(5)$
weights of the {\bf 5} representation. 
The unit cell of the lattice is a compound of two 4-simplex as in 
the ${\bf 5}$ and  
$\mybar {\bf 5}$ representations of $SU(5)$
\footnote{ Three   dimensional counterpart is $A_3^*$ lattice, the body
  centered cubic lattice.  The unit cell  should be 
regarded as a compound ${\bf 4}$ and   $\mybar {\bf 4}$ 
representations of $SU(4)$. 
The  ${\bf 4}$ ($\mybar {\bf 4}$)  is generated by four, three
  dimensional vectors,  $\bfe_m$ ($- \bfe_m$) with $m=1, \ldots 4$,  
 which can be obtained by 
removing the fourth component from  \Eq{latvec}, i.e, by 
dimensional reduction. The compound of two tetrahedron, (the  
${\bf 4}$ and   $\mybar {\bf 4}$) is the famous {\it Stella Octangula} of
Kepler, the simplest polyhedral compound in three dimensions.  The cube
arises  as the convex hull of this object.    
Two  dimensional counterpart is $A_2^*$ lattice, triangular  
  lattice. It  can be thought as  ${\bf 3}$ and   ${\mybar {\bf 3}}$
  representation of $SU(3)$. The  ${\bf 3}$ ($\mybar {\bf 3}$)  is 
generated by three, two
  dimensional vectors,  $\bfe_m$ ($- \bfe_m$) with $m=1, \ldots 3$,  
 which can be obtained by 
removing the third and the fourth component from  \Eq{latvec}.  The $A_4^*$
lattice can be visualized similarly.}.

The matter 
content of $A_4^*$ lattice theory is most easily described in terms 
of representations of $SU(5)$. The ten bosonic degrees of freedom are labeled
as $z^m \oplus \mybar z_m= {\bf 5} \oplus \mybar  {\bf 5} $,  and the sixteen 
fermions are presented as 
$ \lambda \oplus \psi^{m} \oplus  \xi_{mn}=  {\bf 1} \oplus 
{\bf 5} \oplus \mybar  {\bf 10}$.   
The $z^m, \psi^m $ fields reside on the links connecting the center of a
4-simplex to its five corners, which are  labeled by $\bfe_m$. 
The $\mybar z_m$ reside on the links along  $-\bfe_m$. 
 The ten fermions $\xi_{mn}$,
  and ten composite antiholomorphic bosonic fields 
$\mybar E_{mn}= [\mybar z_m, \mybar z_n ] $  reside  on 
 $-\bfe_m -\bfe_n $ directed toward the ten sides of the 4-simplex  and 
finally the  singlet $\lambda$ resides on the site.  
For more details on the $A_4^*$  lattice, see ref.\cite{Kaplan:2005ta}.   
The point group symmetry of the  action  is  permutation group $S_5$, 
the Weyl group of $SU(5)$. Notice that  inversion is not a symmetry, since 
there is no   $\mybar {\bf 5}$ representation in fermionic sector.

Let us reexpress the $A_4^*$ lattice  action for $\CN=4$ theory as 
 a sum of $Q$-exact $\CL_1$ and $\CL_2$,  and $Q$-closed  $\CL_3$ terms. 
Because of symmetry reasons 
and to ease the comparison   with   \Eq{LT4},  
we present  it as $g^2 \CL= \CL_1 + \CL_2 + \CL_3$  where 
 \beq 
 &&  \CL_1 =\sum_{\bfn} 
 Q  \Tr  \lambda_\bfn ( \half id_\bfn + (\mybar z_{m,\bfn -\bfe_m}
z^m_{\bfn -\bfe_m} - z^m_{\bfn} \mybar z_{m,\bfn}))
   \cr
&&  \CL_2 = \sum_{\bfn}  Q  \Tr  \half  {\xi}_{mn, \bfn} 
(z^m_{\bfn} z^n_{\bfn +\bfe_m} -z^n_{\bfn} z^m_{\bfn +\bfe_n} )      \cr
&&  \CL_3 = \sum_{\bfn}  \textstyle{\frac{\sqrt2}{8}} \epsilon^{mnpqr} 
\Tr \xi_{mn, \bfn} ( \mybar z_{p, \bfn -\bfe_p} 
\xi_{qr,\bfn + \bfe_m +  \bfe_n } - \xi_{qr,\bfn - \bfe_q -  \bfe_r }
\mybar z_{p, \bfn + \bfe_m +  \bfe_n } )   
\eqn{A4star}.
\eeq
The supersymmetry transformations  of the lattice fields are given by 
\begin{eqnarray}
&&Q \lambda_{\bfn} =  -id_{\bfn},   \qquad    Q d_{\bfn} = 0 \cr 
&& Q z^{m}_{\bfn} =   \sqrt 2 \, \, \psi^{m}_{\bfn} , \qquad  
Q \psi^{m}_{\bfn} =0 \cr
&& Q \mybar z_{m, {\bfn}} = 0  \cr
&&Q \xi_{mn, \bfn} = -2(\mybar z_{m,\bfn + \bfe_n}  \mybar z_{n,\bfn} - 
 \mybar z_{n, \bfn + \bfe_m}  \mybar z_{n,\bfn} )
\eqn{Qlatticeoffshell}.
\end{eqnarray}
Clearly, the $S_5$ singlet supersymmetry  $Q$ is nilpotent, $Q^2\cdot=0$. In
the rest of this section, we show  the transmutation of 
action \Eq{A4star} into \Eq{LT4} by using the representation theory of
$S_5$. To reduce the clutter, we suppress the lattice indices which transform
in an obvious way under the point group symmetry. 

The physical point group symmetry of the lattice is isomorphic to permutation
group $S_5$.
The character table and conjugacy classes of $S_5$ are 
given in  Table \ref{tab:S5}. 
%%%%%%%%%%%%%%%%%%%%%%%%%%%%%%%
%%%%%%%%%%%%%%%%%%%%%%%%%%%%%%%%%%%
\setlength{\extrarowheight}{-3pt}
\begin{table}[t]
\centerline{
\begin{tabular}
{|c|c|c|c|c|c|c|c|} \hline
classes: & (1)  & (12) & (123) & (1234)&(12345) & (12)(34) & (12)(345)\\ \hline
sizes:   & 1  & 10   &  20   & 30    &  24    &  15      & 20       \\ \hline
$\wchi_1$ & 1  &   1  &   1   &  1    &   1    &    1     &   1      \\ % \hline
$\wchi_2$ & 1  &  -1  &   1   & -1    &   1    &    1     &  -1      \\ %\hline
$\wchi_3$ & 4  &   2  &   1   &  0    &  -1    &    0     &  -1      \\ %\hline
$\wchi_4$ & 4  &  -2  &   1   &  0    &  -1    &    0     &   1      \\ %\hline
$\wchi_5$ & 5  &  -1  &  -1   &  1    &   0    &    1     &  -1      \\ %\hline
$\wchi_6$ & 5  &   1  &  -1   & -1    &   0    &    1     &   1      \\ %\hline
$\wchi_7$ & 6  &   0  &   0   &  0    &   1    &    -2     &   0     \\ \hline
\end{tabular} }
\caption{\sl The character table of $S_5$, the point symmetry group of
  $A_4^*$ lattice. The even permutations are spacetime rotations, the odd 
permutations involves parity operations and hence improper rotations.  
\label{tab:S5}}
\end{table}
%%%%%%%%%%%%%%%%%%%%%%%%%%%%%%%%%%%
%%%%%%%%%%%%%%%%%%%%%%%%%%%%%%%%%%%
The group has $5!=120$ elements and seven conjugacy classes
shown in Table.\ref{tab:S5}.  The symmetry of the lattice action is 
composed of the elements of $S_5$. It is easy to show that even permutations
with determinant   one (the $\wchi_2$ representation) 
are pure rotational symmetries of the action.  
We see from Table.\ref{tab:S5} that the odd 
permutations has determinant 
minus one  (the $\wchi_2$ representation), and are not proper elements of 
$SO(4)'$.  Hence, we consider  $A_5$,  the  rotation subgroup of $S_5$, 
also called alternating group of degree five. 
The $A_5$  is the discrete subgroup of   proper rotations $SO(4)'$
$$
A_5=  S_5/Z_2 \in \; SO(4)'
$$
and we classify fields under $A_5$. 
Also, as noted in \cite{Kaplan:2005ta}, the
odd permutations are symmetries if  accompanied with a fermion phase
redefinitions $\xi\rightarrow i\xi$,   $\psi\rightarrow -i\psi$, and 
 $\lambda \rightarrow i \lambda$. The odd permutations  do not commute with
 supersymmetry  as the field redefinition  treats  the components of a
 supermultiplet differently. 

To classify fields  under $A_5$,  we consider the 
group action   from each of the five conjugacy  classes. The
character table of $A_5$ can be deduced from $S_5$ 
\footnote{
The conjugacy classes of $A_5$ can easily be 
read off from the  character table and conjugacy classes of $S_5$. 
The character table of $S_5$ is given in Table.\ref{tab:S5}. The conjugacy
classes of $S_5$ are the physical symmetries of the 4-simplex.  The conjugacy
classes (which is formed of only even permutations) and the 
sizes of representations of $A_5$ are given in   Table.\ref{tab:A5}.
Notice that in $A_5$, the 5-cycles   splits into  two types.
$(12345)$ and $(21345)$. It is easy to see that only the odd permutations 
(which are absent in $A_5$, but present in $S_5$) can take an element of one
conjugacy class to the other. Hence there are two distinct  conjugacy classes
for five-cycles in $A_5$.

The characters for $A_5$ can also be deduced from  the ones of $S_5$. Since 
the odd permutations are absent in  $A_5$, the sign representation of $S_5$ 
reduce to trivial representation  $\wchi_2|_{A_5}= \chi_1$.  Also, noticing
the relations  $ \wchi_3 \wchi_2 = \wchi_4,  \;   
\wchi_5 \wchi_2 = \wchi_6,  \; \wchi_7 \wchi_2= \wchi_7 $,  
we see that  $\wchi_3|_{A_5} =\wchi_4|_{A_5} = \chi_2$,  
$\wchi_5|_{A_5} =\wchi_6|_{A_5} = \chi_3$. Finally, the  $\wchi_7$ is reducible 
in $A_5$. From  the relation, $\wchi_7 \wchi_2= \wchi_7$, we see that 
$\wchi_7$ is zero for all odd permutations. It splits as  $\wchi_7|_{A_5}= 
\chi_4 + \chi_5$ into two three dimensional representations.
As a physical consequence, 
unlike $S_5$, the $A_5$ can not distinguish a scalar from
pseudo-scalar and a vector from a pseudo-vector. 
}  
and is given in    Table.\ref{tab:A5}.
%%%%%%%%%%%%%%%%%%%%%%%%%%%%%%%
%%%%%%%%%%%%%%%%%%%%%%%%%%%%%%%%%%%
\setlength{\extrarowheight}{3pt}
\begin{table}[t]
\centerline{
\begin{tabular}
{|c|c|c|c|c|c|c|c|} \hline
classes: & (1)    & (123) &(12345) & (21345)  &  (12)(34) \\ \hline
sizes:   & 1     &  20       &  12 & 12    & 15       \\  \hline
$\chi_1$ & 1  &   1  &   1   &  1    &   1         \\  %\hline
$\chi_2$ & 4   &   1       &  -1    &    -1     &  0      \\  %\hline
$\chi_3$ & 5  &   -1  &  0 & 0    &   1       \\  %\hline
$\chi_4$ & 3  &   0  &   $\rm{\frac{(1+\sqrt5)}{2}}$   &
$\rm{\frac{(1-\sqrt5)}{2}}$  &   -1       
\\ %\hline
$\chi_5$ & 3  &   0  &    $\rm{\frac{(1-\sqrt5)}{2}}$  &  
$\rm{\frac{(1+\sqrt5)}{2}}$ &   -1       
\\ \hline
\end{tabular} }
\caption{\sl The character table of $A_5$, the rotation  subgroup of $S_5$.
 The pure rotational symmetries of  $A_4^{*}$ lattice.
\label{tab:A5}}
\end{table}
%%%%%%%%%%%%%%%%%%%%%%%%%%%%%%%%%%%
%%%%%%%%%%%%%%%%%%%%%%%%%%%%%%%%%%   
By choosing a representative  from each conjugacy  class,  
we calculate the character of the corresponding group element.  
In Table.\ref{tab:op}, we show how 
an element from each class acts on the link field and calculate 
%%%%%%%%%%%%%%%%%%%%%%%%%%%%%%%
%%%%%%%%%%%%%%%%%%%%%%%%%%%%%%%%%%%
\setlength{\extrarowheight}{-2pt}
\begin{table}[t]
\centerline{
\begin{tabular}
{|l|c|c|} \hline
Operation   & $ z^1, z^2, z^3, z^4, z^5 $ & $\chi(g_{(rep)})$ 
\\ \hline
(1)    & $ z^1, z^2, z^3, z^4, z^5 $ & 5   
\\ \hline
(123)  & $ z^3, z^1, z^2, z^4, z^5 $  & 2 
\\ \hline 
(12345)    & $z^5, z^1, z^2, z^3, z^4  $  & 0
 \\ \hline
(13452)    & $ z^2, z^5, z^1, z^3, z^4  $  & 0
 \\ \hline
(12)(34)   & $ z^2, z^1, z^4, z^3, z^5 $  &1
\\ \hline
\end{tabular}}
\caption{\sl A representative of each conjugacy class and their action  
on the site and link fields are shown in the table. The five link fermions 
$\psi^m$ transform in the same way with $z^m$. The transformation of 
 ten fermions $\xi_{mn}$ can be deduced from the antisymmetric product
 representation of $\mybar z_m$ with itself. 
\label{tab:op}}
\end{table}
%%%%%%%%%%%%%%%%%%%%%%%%%%%%%%%%%%%
the character $\chi(g)= \Tr( O(g))$, where $g$ is a 
representative of each class and $O$ is a 
matrix representation of the operation.  
Since the  character is a class function, it is independent of 
representative.  
A  simple calculation for all the lattice  fields yields 
\beq
&&\chi(\lambda)=\chi(d)=  (1,1,1,1,1) \sim \chi_1  \cr
&&\chi(z^m)=  \chi(\psi^m )= \chi(\mybar z_m)= (5,2,0,0,1)   
\sim  \chi_2 \oplus \chi_1  \cr
&&\chi(\xi_{mn})=\chi(\mybar E_{mn})=  (10, 1,  0, 0, -1,) \sim \chi_2 \oplus 
\chi_4 \oplus \chi_5 \, . 
\eqn{A5splitting}
\eeq
By inspecting the character table, 
we observe that the link fields 
are indeed in  reducible 
representations of the point  group symmetry $A_5$.  
The  site multiplet is in trivial   representation.  
\beq
\chi(\lambda)= \chi(d)= \chi_1 \, .
\eeq
The five link fields splits as 
\beq
\chi(z^m)= \chi(\psi^m)=  \chi(\mybar z_m)=   \chi_2  \oplus \chi_1, 
\eeq
into a singlet representation and a four dimensional
representation. Similarly, the ten  
 fermions  $\xi_{mn}$ decompose into a four and two three dimensional 
representations as
\beq
\chi(\xi_{mn})= \chi({\mybar E}_{mn})=  \chi_2 \oplus \chi_4 \oplus \chi_5
\eeq 
One can also show that the product of fields which make the 
${E}^{mn}= [z^m, z^n]$ function   transforms as
\beq
\chi({ E}^{mn})= 
[\chi( z^m) \otimes \chi (z^n)]_{A.S.} =  
\chi_2 \oplus \chi_4 \oplus \chi_5
\eeq
Notice that the splitting of $\chi(\xi_{mn})$ and $\chi({ E}^{mn})$  in $S_5$  
is $\wchi_3 \oplus \wchi_7$ into a four and six dimensional representation. 
Under $A_5$,  $\wchi_7|_{A_5}=  \chi_4 \oplus \chi_5$ 
splits further because of lower symmetry.    
We observe that the elementary fermionic and bosonic degrees of freedom 
split into  irreducible representations as
\beq 
&& {\rm  fermions}
\rightarrow 2 \chi_1 \oplus  2 \chi_2 \oplus  
[\chi_4 \oplus \chi_5] = 2(1) \oplus 2(4) \oplus [3\oplus 3'] \cr
&& {\rm bosons}\rightarrow 2 \chi_1 \oplus  2 \chi_2 
= 2(1) \oplus 2(4) \, ,
%\cr && {\rm composite} \; E_{mn} \; {\rm function} \rightarrow 
%\chi_2 \oplus [\chi_4 \oplus \chi_5] = 4 \oplus  [3\oplus 3]
\eqn{split}
\eeq
where  the dimension of the corresponding irreducible representations is
written explicitly. 
This is indeed the branching of fermions and bosons under the twisted 
Lorentz symmetry $SO(4)'$ discussed in  section.\ref{sec:marcus}. 
It is easy to identify the continuum fields (which transform under the
irreducible representations of twisted rotation symmetry $SO(4)'$) with the 
irreducible representation of the discrete rotations on the lattice.
 The two  scalar fermions of the continuum theory are associated with the 
two singlet ($\chi_1$)  fermions  on the lattice.  Similarly,
the vector and three form of the continuum are the two   four dimensional 
$\chi_2$ representation. Finally, the six fermions (in two index
antisymmetric representation) of the continuum theory $6$ of $SO(4)'$ 
 reside in    the two three dimensional representation  
$\chi_4 \oplus  \chi_5$ of the $A_5$, or better in  $\wchi_7$ of $S_5$.
The self-dual and antiselfdual splitting of $6$ into  
$[(3,1) \oplus (1,3)]$  representations takes place in the spin group of 
$SO(4)'$ and these two three dimensional representation is  not related 
to  $\chi_4 \oplus  \chi_5$ of $A_5$.
 The bosonic degrees of freedom   work similarly. 

How can we compute these irreducible representations explicitly?  For
example, for link fields $z^m$, what does the splitting  
$\chi_2 \oplus \chi_1$ mean? Recall that under a group operation (see 
Table.\ref{tab:op}), $z^m \rightarrow O^{mn}(g)z^n$. Dropping all the
indices, $z'= Oz$. The fact that the group action on the link field is
reducible means there is a similarity transformation which takes all of the 
$O(g)$ into a block diagonal form. In this case, two blocks have  sizes 
$1\times 1$ and  $4\times 4$.  This naturally splits the $z_m$ vector space 
into two components of size one and four, which never mixes under group
action. It is easy to guess the singlet representation: it is 
$\frac{1}{\sqrt 5}\sum_{m=1}^{5}z^m $.  Now, let us introduce a $5 \times 5$
orthogonal  matrix $\CE$ that block-diagonalizes $O(g)$ for all $g\in A_5$. 
Then  we have  $(\CE^{-1} z')=(\CE^{-1} O \CE) (\CE^{-1} z)$. \
A little bit work shows
that  the $\CE$ matrix can be expressed in terms of components of the
basis  vectors $\bfe_m$ \footnote{This relation is  true for all $A_d^{*}$
 lattices. In arbitrary $d$, we have $d+1$ linearly dependent $d$-dimensional 
vectors satisfying  $\bfe_m. \bfe_n= \delta_{mn}- \frac{1}{d+1}$ and 
$\sum_{m=1}^{d+1} \bfe_m =0$. The block diagonalization  matrix $\CE$ takes
the form
$ 
\CE_{m\mu}= (\bfe_m)_{\mu}, \; \CE_{m, d+1}= \frac{1}{\sqrt {d+1}}
$. The matrix $\CE$ connects the irreducible representations of $S_{d+1}$ to
the ones of $SO(d)'$.}
\beq
\CE_{m\mu}= (\bfe_m)_{\mu},  \qquad \CE_{m5}= \frac{1}{\sqrt5} \, .
\eeq
The matrix $\CE_{m n}$ forms a bridge between the irreducible 
representation of  $A_5$ and the  representations
 of the twisted Lorentz  group $SO(4)'$. 
Thus, we obtain the following relations dictated by  symmetry arguments:
\beq
 \CE_{m\mu } z^m_\bfn = z^{\mu} (x) \qquad 
\CE_{ m5} z^{m}_\bfn = \epsilon_{\mu \nu \rho \sigma} z^{\mu\nu \rho \sigma}(x)/24 
\eeq
For  fermions $\xi_{mn}$, it is easy to show that  the continuum 
fields are $ \xi_{mn, \bfn} \CE_{m\mu}\CE_{n\nu}= \xi_{\mu\nu}(x)$ and  
$\xi_{mn,\bfn } \CE_{m\mu}\CE_{n5}= \epsilon_{\mu
  \nu \rho \sigma}\xi^{\nu \rho \sigma}(x)/6$.  
Similarly, the antiholomorphic function $\mybar E_{mn, \bfn}$ splits into 
$\mybar \CF_{\mu \nu}(x)$ and  $\mybar \CD_{\mu} \mybar z^{\mu \nu \rho
  \sigma}(x)$. 
This completes 
our discussion of the relation between the $A_4^{*}$ lattice action and 
twisted theory \Eq{LT4}.  The continuum limit of the 
Lagrangian \Eq{A4star} at tree level   reproduce  the twisted theory  
\Eq{LT4}. 

\subsection{Connection  with Catterall's formulation} 
\label{sec:othertwist}
Another recent proposal for lattice regularization of $\CN=4$ SYM theory 
had been introduced by S. Catterall. In this section, we want to briefly 
mention  the   relation  between the  two approaches.  In fact, upon realizing
the  fact that the orbifold lattice produces Marcus's twist,   this 
is an obvious task. Catterall already  provides the mapping between the 
Lagrangian  \Eq{LT4}  and the action he employs in latticization
\cite{Catterall:2005fd}. Here, we construct the relation only in the 
sense of supersymmetry subalgebras 
that are  manifest in these two lattice constructions. 
 
Let us rename the  twist introduced in the previous
section as an A-type twist. In fact, there is another scalar supersymmetry, 
associated with Poincar{\'e}  dual of the 4-form Grassmann $* \psi^{(4)}$.
We could  have chosen  $\mybar Q= *Q^{(4)}= \frac{1}{4!}
\epsilon_{\mu \nu \rho\sigma}Q^{\mu \nu \rho \sigma}$   
as the manifest scalar supersymmetry.  We call this B-type supersymmetry. 
To make the comparison with the Catterall lattice formulation, it is 
convenient to dualize the 3-form and 4-forms fields to vectors and scalars  
respectively.
\beq
\frac{1}{4!} \epsilon^{\mu \nu \rho \sigma}  (z_{\mu \nu \rho \sigma}, 
\psi_{\mu \nu \rho \sigma}) = (z, \psi)
\qquad  \frac{1}{4!} \epsilon_{\mu \nu \rho \sigma}  \mybar z^{\mu \nu \rho
  \sigma}= \mybar z, \qquad \frac{1}{3!} \epsilon_{\mu \nu \rho \sigma}  
 \xi^{\nu \rho
  \sigma}= \chi_{\mu},
\eeq
The continuum on-shell A-type supersymmetry transformation  are given by 
\footnote{ The transformations in  \Eq{QAonshell},   
\Eq{QBonshell} and \Eq{QAB}
can easily be read off of the transformation given in Eq.(4.9) of 
ref.\cite{Kaplan:2005ta} by using the substitution $z^{\mu}\rightarrow 
 \CD^{\mu}/ \sqrt{2} $ and  $\mybar z_{\mu}\rightarrow  \mybar \CD_{\mu}/ 
\sqrt2 $.}

\begin{eqnarray}
&&Q \lambda =  -  (\,   [\mybar z,  z] + \half [\mybar \CD_{\mu}, \CD^{\mu}]
   \, ) \cr
&& Q z^{\mu} =   \sqrt 2 \, \, \psi^{\mu}, \qquad  Q \psi^{\mu}=0 \cr
&& Q \mybar z_{\mu} = 0  \cr
&&Q \xi_{\mu\nu} = -i \mybar \CF_{\mu \nu} \cr
&&Q \chi_{\mu} =  \sqrt 2 \, \mybar \CD_{\mu}  \mybar z \cr
&&Q z = \sqrt{2} \psi, \qquad 
Q \psi = 0    \cr 
&&Q \mybar z =0   
\eqn{QAonshell}
\end{eqnarray}
and similarly  the on-shell B-type transformations are 
\beq 
\qquad &&\mybar Q \lambda = 0 \cr
&& \mybar Q z^{\mu} =  0 , \qquad  \mybar  Q \psi^{\mu}= \sqrt 2 {\CD}^{\mu}
\mybar \phi \cr 
&& \mybar Q \mybar z_{\mu} = - \sqrt 2 \chi_{\mu}, \cr 
&&\mybar Q \xi_{\mu\nu} =-i (\half \epsilon_{\mu \nu \rho \sigma}) 
\CF^{\rho  \sigma} \cr
&&
\mybar Q \chi_{\mu}=0  \cr
&&\mybar Q z = \sqrt{2} \lambda, 
\qquad 
\mybar Q \psi = ( \half [\mybar \CD_{\mu}, \CD^{\mu}]  -[ \mybar z, z] )\cr 
&&\mybar Q \mybar z =0  
\eqn{QBonshell}  
\end{eqnarray}
Notice that both $Q$ and $\mybar Q$ are nilpotent: $Q^2=0,\;\;
 {\mybar Q}^2=0$, 
up to the use of equation 
of motion. As we have seen in the previous section,  an off-shell completion 
is possible by introducing an auxiliary field $d$.  
A linear combination of  A and B-type scalar supersymmetries is the exact 
manifest supersymmetry that is utilized in  Catterall's formulation.  
Since the $U(1)$  charges of these two supercharges are equal, we can add
them without upsetting this symmetry.   
Using the supercharge 
\beq
\wQ= \frac{Q^{(0)} + *Q^{(4)}}{\sqrt2}= \frac{Q + \mybar Q } {\sqrt2} 
\eqn{QAB}
\eeq
we observe that  the off-shell $\wQ$-action on fields are given by 
\begin{eqnarray}
 \wQ z^\mu &=&  \, \, \psi^\mu, \qquad   \qquad
\wQ \psi^\mu = \, \CD^{\mu} \mybar z, \cr
 \wQ \mybar z_{\mu} &=& -   \, \, \chi_{\mu}, \qquad \;\;\;
 \wQ \chi_{\mu} = - \,  \mybar \CD_{\mu} \mybar z  \cr 
\wQ z &=&  (\psi  + \lambda), \qquad \wQ (\psi + \lambda)
 =  -  \sqrt 2 \,  [\mybar z, z] \cr
\wQ \mybar z &=& 0,   \qquad \qquad \qquad
\wQ (\psi - \lambda) =  {\textstyle \frac{1}{\sqrt2}}
[\mybar \CD_{\mu}, \CD^{\mu}]  \cr
\wQ \xi_{\mu\nu} &=&     {\textstyle \frac{-i}{\sqrt2}}
( \mybar \CF_{\mu \nu}  + \half  \epsilon_{\mu\nu\rho\sigma}  \CF^{\rho
  \sigma})
\eqn{Qtwisted}
\end{eqnarray}
The $\wQ$ transformation satisfies 
\beq 
\wQ^2 \cdot  =  \delta_{\mybar z}\; \cdot  \; ,
\eqn{QAB2}
\eeq
 which can seen by using equations of motion. 
Here,  $\delta_{\mybar z}$ is a field dependent infinitesimal 
gauge transformation. 
Notice that  $\CQ$ is not exactly nilpotent, but nilpotent up to  a gauge 
rotation. Catterall employs  \Eq{QAB} as the exact manifest supersymmetry 
on the lattice. Naturally,  continuum actions in terms of propagating degrees 
of freedom can be easily mapped into each other.  However, the number of 
bosonic off-shell degrees of freedom are not same in the two formulation.  
This can be understood by working the  off-shell completion of the  
supersymmetry  algebra  \Eq{QAB}. It  is different from 
\Eq{QAoffshell} and necessitates  introducing a two form auxiliary  field.  
For the details of this construction, see
\cite{Catterall:2005fd}. I do  not know the precise relation with the 
formulation of Sugino \cite{Sugino:2004uv}, 
but similar considerations may hold.  However, I want to comment on the merit
of having more than one formulation in a somewhat  idiosyncratic  way, 
by using reasonings  from the calculations of topological correlators  
in  the  continuum formulation.

\subsection{The fermion sign problem and topological  correlators}
The  extended supersymmetric gauge theories shown in
Table.\ref{tab:tab1}   in general have a fermion sign
problem even in continuum. 
In the case of $\CN=4$ SYM theory, the source of the sign problem
can be traced to the Yukawa interactions, and therefore to 
nonvanishing  field configurations of   scalars. 
Conversely,  in $\CN=1$ SYM in $d=4$ dimensions, a theory without scalars, 
the positivity of the fermion Pfaffian  can be proven. 
Here,  I will argue that for a  very restricted class of 
observables, the fermion sign problem  should not be a problem. 
Similar considerations may hold in  some  lattice formulations as well.
Unfortunately, this class is really small and 
the consideration of this  section does not mean much for the full set of
correlators of the physical theory. However, one can also pursue a more 
optimistic complementary logic \cite{  Giedt:2003vy}. 
Since many things are known or conjectured 
about the 
$\CN=4$ or other highly  supersymmetric target theories, this data can be
used to make progress in the understanding of the sign problem. After all,
the sign problem arise because of inadequacy of the path integral, and 
is not a pathology of the theory.  The reason that one can evade sign problem
for topological correlators is a  localization property of the path integral 
that we explain below.  
The  ideas in this section borrows directly 
from the Witten's classic construction of topological field theory
\cite{Witten:1988ze} and adopts  the arguments there to  the $\CN=4$ 
SYM  theory.  

The transformations  \Eq{Qtwisted} look rather similar to the ones 
introduced  by Witten in the study of the Donaldson theory 
\cite{Witten:1988ze}. 
Indeed, the supersymmetry algebras are identical, $\wQ^2= \delta_{\mybar z}$.  
The difference is in the  field content. 
  Witten considers   the twist of $\CN=2$ theory, an asymptotically free 
theory  in $d=4$ dimensions,
and addresses questions about the  topological correlators (in the sense of 
$\tilde Q$) in the twisted theory. For the calculation of 
topological correlators, to regard  $\tilde Q$ as a BRST and to 
make the theory 
truly topological is a matter of preference. One can consider the physical 
theory and still calculate correlators in the  topological  sector. 
Simplest examples of this type is the  supersymmetric quantum mechanics 
with discrete spectrum. (For continuous spectrum, supersymmetry does not imply 
the equality of density of states in the bosonic and fermionic sector and the 
following statements needs  refinement.) 
 For  example, in the calculation of topological 
partition function, 
(Witten index), one can  sum over all states in the Hilbert space. 
There is  an exact  cancellation between paired bosonic and fermionic states 
with  nonzero  energy, and  hence only zero energy states contribute. 
 Alternatively,  one can 
declare the theory topological, and the physical states  ($Q$-cohomology 
group) of the topological 
theory   are just  the quantum ground states of the full theory. 
The rest of the   Hilbert space  is  redundant in the sense of BRST,  and 
the partition function only receives contribution from quantum ground states. 
In the language of path integrals, this translates to localization of 
appropriately defined  correlators 
to the the   fixed point of the $Q$-action in the  supersymmetry 
transformations, such as the ones \Eq{QAonshell} and \Eq{Qtwisted}. 
Therefore, there are  stronger techniques to calculate topological 
correlators. See for example for a review \cite{Cordes:1994fc}.

Marcus  shows that the fixed point of \Eq{QAonshell} is the space of
complexified flat connections. 
He also argues that the theory reduces to
Donaldson-Witten theory if one demands a reality  condition and use the 
 $\tilde Q$ in  \Eq{Qtwisted} \cite{Marcus:1995mq}. 
\footnote{ This reality condition is  not compatible 
with the gauge invariance on the lattice construction.  
If it were possible to implement 
this condition,  this   would yield a lattice formulation of $\CN=2$ SYM
   theory in four dimensions and would be remarkable.}
Then, in  \Eq{Qtwisted}, both field strengths 
reduce to the usual field strength, i.e, $\mybar \CF_{\mu \nu}= 
\CF_{\mu \nu}= F_{\mu \nu}$. The vanishing condition of the final equation in
that case becomes the instanton equation  $F_{\mu \nu}+ \half \epsilon_{\mu \nu
  \rho \sigma}F^{\rho \sigma}=0$ and  the  path integral 
 can be expanded around the  instantons. 
Here, we do not wish to make such an assumption and just consider the theory
as it is. This gives a  complex version of instanton equations 
which relates the holomorphic field strength to the dual of the
anti-holomorphic field strength: 
\beq 
\mybar \CF^{(2)} + *\CF^{(2)}=0, \qquad  
\mybar \CF_{\mu \nu} +  \half  \epsilon_{\mu \nu
  \rho \sigma} \CF^{\rho \sigma}=0
\eqn{cie}
\eeq
or equivalently using \Eq{strength}, we can split it into its hermitian 
and antihermitian parts. In this case, the equation takes the form:
\beq
&& F_{\mu\nu} -i [S_{\mu}, S_{\nu}] +   \half  \epsilon_{\mu \nu
  \rho \sigma}  (F^{\rho \sigma} -i [S^{\rho}, S^{\sigma}]) = 0  \cr
&&  (D_{\mu}S_{\nu}- D_{\nu}S_{\mu})-     \half  \epsilon_{\mu \nu
  \rho \sigma}  (D^{\rho}S^{\sigma}- D^{\sigma}S^{\rho})=0 
\eqn{cie2}
\eeq
I do not know the full set of solutions to these equations. However, 
it seems rather plausible that the moduli space 
(as in Donaldson-Witten theory) is just  isolated instantons under 
circumstances  analyzed in 
 \cite{Witten:1988ze}.  Then, by using the weak coupling limit of the 
theory and by exploiting the coupling constant  independence  of the  
partition function, one can calculate certain observables. It seems sufficient
to keep the quadratic part of the Lagrangian (owing to weak
coupling) and  benefit from the steepest decent techniques.    
If all this holds, 
then the fermionic determinant  around such instanton configurations should 
be real, and by supersymmetry should be related to the bosonic determinant. 
This  simply follows from the equality of nonzero 
eigenvalues of the bosonic and fermionic quantum fluctuations 
around the instanton background.
For example, under circumstances where the dimension of the instanton moduli 
space is zero, and hence there are no fermionic zero modes, 
the partition function  of the theory should be a 
topological  invariant  \cite{Witten:1988ze}
and should be calculable without any reference to 
fermion sign problem.  Similar considerations also hold for other 
topological correlators.
The main point is that the fixed points of  some   $Q$-actions may lead to
the finite action field configurations  which admits the  saddle-point 
approximations.  In the case where  the observables are independent of 
coupling  constants, the partition function localizes to these fixed points 
and hence dominates the path integral. Under such circumstances, one can
evade fermion sign problem. Also see \cite{Cordes:1994fc} about
localization. 
   
\section{ The Blau-Thompson twists and three  dimensional lattices}
In this section, we show that  the orbifold lattice action of the 
three  dimensional theories  produce the Blau-Thompson type twists 
\cite{Blau:1996bx}. 
 In each case, 
we will see that the point group symmetry of the lattice action  
enhances to  the  twisted rotation group $SO(d)'$ in the continuum. 
We will also observe that 
a continuous $R$-symmetry which has  the same rank  as the $R$-symmetry 
of the continuum twisted  theory is exactly realized 
on the cubic orbifold lattices. The features of these lattices in the 
sense of representation theory follows very similar pattern to our discussion
in the previous section. Namely, there are always spacetime scalars  in
vector representation of the twisted rotation group and hence lattice, and
the double valued spinor representation of the continuum theory are always
associated with the single valued representations of the orbifold lattice
theories. 
Since all the tools that we need to use are developed   
in the  previous section, our presentation will be brief and
will emphasize symmetries rather than technical details.  

\subsection{The  $\CN=4$ SYM in $d=3$}
\label{sec:blau}
The $\CN=4$ SYM theory in three dimensions possess a global  
$G=SU(2)_E\times SU(2)_{R_1} \times SU(2)_{R_2} $  where $SU(2)_E \sim SO(3)_E$
 is the Euclidean Lorentz symmetry and   $ SU(2)_{R_1} \times SU(2)_{R_2} $
 is the $R$-symmetry of the theory. 
To construct the   Blau-Thompson twist \cite{Blau:1996bx}, 
we take the diagonal subgroup of  
the spacetime $SO(3)\sim SU(2)_E$ and  $SU(2)_{R_1}$.
The twisted theory possess an   $SO(3)' \times SU(2)_{R_2}$  symmetry.

Under $G$, the vector boson, scalars and fermions  transform  as $(3,1,1)$,
 $(1,3,1)$,  $(2,2,2)$.  In the twisted theory, the the gauge bosons and 
scalars are on the same footing and they  transform as    $(3,1)$.   
The  fermions splits as $(3,2)\oplus (1,2)$,  both of which are
 doublets  under $SU(2)_{R_2}$. However, our lattice only respects the
$U(1)$ subgroup of  the   $SU(2)_{R_2}$ and the full $SU(2)_{R_2}$ only
emerges  in the continuum. Therefore, we will express the continuum action 
with manifest 
 $ G'=SO(3)' \times  U(1)$ symmetry.
The  fermions and bosons  under $G'$  transform as 
$${\rm fermions}\rightarrow 1_{\frac{1}{2}}\oplus 3_{-\frac{1}{2}} \oplus  
3_{\frac{1}{2}} \oplus 1_{-\frac{1}{2}}, \qquad  {\rm bosons}\rightarrow  
3_{0} \oplus 3_{0} .
$$  
We label the fermions 
as $(\lambda, \psi^{\mu}, \xi_{\mu \nu}, \xi^{\mu \nu \rho})$. The action of
the twisted theory is 
\beq
\CL =&& \frac{1}{g^2} \Tr \Big[ Q   \left( \lambda ( \half id  +  \half 
[\mybar {\cal D}_{\mu},  \CD^{\mu}] )
+ {\textstyle \frac{i}{4}} \xi_{\mu\nu} {\cal F}_{\mu\nu} 
\right) + \half  \xi^{\mu\nu \rho} \mybar {\cal D}_{\mu} \xi_{\nu \rho}  \Big] \cr
 = &&   \frac{1}{g^2}  \Tr  \Big[ \eight  ([\mybar {\cal D}_{\mu} , 
 {\cal D}^{\mu}])^2 + \fourth  |{\CF}_{\mu\nu}|^2 + 
 \lambda  \mybar {\cal D}_{\mu} \psi^{\mu} +  
 \xi_{\mu\nu}  {\cal D}^{\mu} \psi^{\nu} 
+ \half  \xi^{\mu\nu \rho} \mybar {\cal D}_{\mu} \xi_{\nu \rho}  \Big]
\eqn{LT3}
\eeq
The off-shell $Q$-transformations are given by  
\beq
&&Q \lambda= -id, \qquad   Q d = 0 \cr
&&Q z^{\mu} = \sqrt 2 \psi^{\mu}, \qquad Q \psi^{\mu}= 0,  \cr
&&Q\mybar z_{\mu}=0   \qquad   \mu=1, \ldots 3    \cr
&&Q \xi_{\mu \nu}= -i \mybar \CF_{\mu \nu} \cr 
&&Q \xi^{\mu \nu \rho }= 0 
\eeq
where $Q^2=0$. 
The action is a sum of a $Q$-exact  and $Q$-closed term. The $Q$-invariance  
of the $Q$-closed term may be seen by the use of  Jacobi identity \Eq{Jacobi}. 

{\bf Cubic Lattice: \;}The three dimensional orbifold lattice action for 
$\CN=4$ $d=3$ theory \cite{Cohen:2003qw}  is a simple latticization of the  
Blau-Thompson twist of the theory.  The lattice possess an $S_3 \ltimes
Z_2$ point group and a continuous  $U(1)$ $R$-symmetry group.  
 Inspecting the fermionic 
degrees of freedom; we observe that the fermions are associated with 
$p$-cells: one site, three links, three faces and one cube.  
\beq
{\rm fermions} \rightarrow 1 \oplus 3 \oplus  3 \oplus  1 \; .
\eeq  
In the continuum, they fill  the  antisymmetric tensor representation of 
$SO(3)'$.   Similarly, the vector bosons reside  on the links, and 
they distribute as 
\beq
{\rm bosons} \rightarrow 3 \oplus 3\; . 
\eeq   
In the continuum, they form the vector presentation of $SO(3)'$. 
The lattice formulation   along with the details  of the  superfield 
formulation of  the twisted theory is given in \cite{Cohen:2003qw}. 

\subsection{The $\CN=8$ SYM in $d=3$}
The  $\CN=8$ SYM in $d=3$ theory in $d=3$ dimensions  possess a global 
$G = SO(3)_E \times SO(7)_R$ symmetry. Under $G$,  the fields transform
as 
\beq
{\rm fermions} \rightarrow   (2,8), \qquad
{\rm gauge \; boson} \rightarrow   (3,1), \qquad  
{\rm scalars} \rightarrow   (1,7),
\eeq
 In order to construct the Blau-Thompson twist \cite{Blau:1996bx, 
Geyer:2001yc} 
we  decompose the  $R$-symmetry as   
$SO(7)_R \rightarrow SO(3)_{R_1} \times SO(4)_{R_2}
 \sim SU(2)_{R_1} \times (SU(2) \times SU(2))_{R_2}$. 
Under this decomposition, the scalars and fermions 
splits as $7 \rightarrow (3,1, 1) \oplus (1,2,2)$ 
and   $8   \rightarrow  (2, 2,1) \oplus (2, 1, 2)$.
As usual,  we take the  diagonal sum   of the Euclidean rotation group and 
the $R_1$-symmetry group. 

The twisted theory is invariant under 
$G'= SO(3)' \times (SU(2) \times SU(2))_{R_2}$,   and the 
 fields transform under $G'$ as  
 \beq
&&{\rm fermions} \rightarrow   (1, 2,1) \oplus (3,2,1) \oplus (3, 1,2) \oplus 
(1,1,2) \cr 
&&{\rm gauge \; boson} \rightarrow   (3,1,1), \qquad  
{\rm scalars} \rightarrow   (3,1,1) \oplus (1,2,2) \; .
\eeq
As in the case of the $\CN=4$ theory in three dimensions, 
even though $(SU(2)\times SU(2))_{R_2}$ 
is a symmetry of the continuum theory, the orbifold lattice only 
respects an abelian $U(1)\times U(1)$ subgroup. The full non-abelian
symmetry emerges as an accidental symmetry in the continuum.  
The transformation of the fields under  $SO(3)' \times U(1) \times U(1)$ 
may be summarized as 
\beq 
z \oplus \mybar z \rightarrow 1_{1,0} \oplus 1_{-1,0},   \qquad
z^{\mu} \oplus \mybar z_{\mu} \rightarrow 3_{0,0} \oplus 3_{0,0}, \qquad 
z_{\mu \nu \rho} \oplus  \mybar z^{\mu \nu \rho} \rightarrow 1_{0,1} \oplus 1_{0,-1} 
\eeq 
for the ten bosonic degree of freedom and 
\beq 
&&\lambda \oplus  \psi^{\mu}  \oplus \xi_{\mu \nu} \oplus  
\chi^{\mu \nu \rho}  \rightarrow  
1_{\frac{1}{2}, \frac{1}{2}} \oplus 
3_{-\frac{1}{2}, -\frac{1}{2}}   \oplus 
3_{\frac{1}{2}, \frac{1}{2}} \oplus 1_{-\frac{1}{2}, -\frac{1}{2}} \cr 
&& \alpha  \oplus  \chi_{\mu} \oplus \mybar 
\xi^{\mu \nu}   \oplus  \psi_{\mu \nu \rho} 
\rightarrow  
 1_{\frac{1}{2}, -\frac{1}{2}} \oplus
3_{-\frac{1}{2}, \frac{1}{2}} \oplus
3_{\frac{1}{2}, -\frac{1}{2}}
\oplus  1_{-\frac{1}{2}, +\frac{1}{2}} 
\eeq
for the sixteen fermions. 

{\bf Cubic Lattice:}
The three dimensional lattice action for $\CN=8$ $d=3$ theory   
reproduces the Thompson-Blau twist  in the continuum.  The symmetries 
of the cubic lattice action are $S_3\ltimes Z_2 \ltimes Z_2$ point group 
and a $U(1)\times U(1)$ $R$-symmetry group.    
The distribution of the fermions on the lattice follows very similar pattern 
to  the $\CN=4$ theory. Since the number of fermions is doubled with 
respect  to  $\CN=4$,  each $p$-cell accommodates twice as many fermions. 
The fermions distributes  to $p$-cells as 
\beq
{\rm fermions} \rightarrow 2 (1 \oplus 3 \oplus 3 \oplus 1) 
\eeq
on the lattice.   
Similarly, the elementary bosons  reside on the sites, links, and  3-cell
and  they distribute as 
 \beq
{\rm bosons}  \rightarrow 2 (1 \oplus  3 \oplus 1) \; . 
%\qquad {\rm  auxiliary \; field  }  \rightarrow 1
\eeq   
In the continuum, they are scalars, 
vectors and antisymmetric third rank tensors under $SO(3)'$.   

{{$\pmb A_3^*$} {\bf  (bcc) lattice:}}  
In order to see the  Blau-Thompson twist from the body centered 
cubic (bcc) lattice, we follow the strategy of section \ref{sec:A4star}. 
The lattice action possess the octahedral symmetry  $ 
O_h \sim S_4 \ltimes Z_2 $ 
where $S_4$ is the permutation group and $Z_2$ is the inversion group.  
The lattice also has a charge conjugation symmetry. \footnote{The
  octahedral symmetry group $O_h$ may be constructed in two different 
ways.  One is $O_h \sim T_d \ltimes Z_2  $ where  $T_d$ is the symmetry group 
of tetrahedron and the other is  $O_h \sim O \ltimes Z_2  $ where $O$  is 
the rotation subgroup of $O_h$. The  $Z_2$ is inversion. 
In identifying the lattice fields with the  Blau-Thompson twisted version of
the continuum, it is  sufficient to work with $S_4/Z_2 =A_4$. 
Both $A_4$ and $S_4$   respect   holomorphy for bosonic
multiplets. The $S_4$ group actions on lattice fields  turns
bosonic (anti)-chiral supermultiplets into (anti)-chiral ones. However, the
fermionic chiral and anti-chiral multiplets mixes under the $S_4$
action.  The $Z_2$ inversion exchanges chiral and antichiral bosonic 
multiplets.}

We classify fields under the rotational subgroup $A_4$ of the tetrahedral group
$S_4$.\footnote{\label{foot:ham}The same lattice structure also shows up in
  spatial lattice formulation of $d=4$ dimensional $\CN=4$ theory which is
  suitable for a Hamiltonian formulation \cite{Kaplan:2002wv}. The analysis 
of the irreducible representations of the full $S_4\ltimes Z_2 \ltimes Z_2$  
symmetry (the last
$Z_2$ is charge conjugation) should be helpful  to map the correlation
functions   of the continuum  to the ones on the lattice.}  
Therefore,  we consider the group action from each of the 
four  conjugacy classes of $A_4  $ and calculate the characters.  
For the one index link fields, we find that there is an $A_4$-invariant 
subspace as in the case of the four dimensional lattice and 
these link fields are indeed  reducible. The 
two index link fields  are also reducible. 
A simple calculation yields 
\beq
&& \chi(\lambda)= \chi(\alpha)= (1,1,1,1)= \chi_1 \cr
&&\chi(z^m)= \chi(\psi^m)=\chi(\mybar z_m)= \chi(\mybar \psi_m)= 
(4,  1, 1, 0) =  \chi_4 \oplus  \chi_1 \cr 
&& \chi(\xi_{mn})= ( 6, 0, 0, -2) = \chi_4 \oplus  \chi_4 \; ,
\eeq
which is a natural counterpart of the result \Eq{A5splitting}. 
%%%%%%%%%%%%%%%%%%%%%%%%%%%%%%%
%%%%%%%%%%%%%%%%%%%%%%%%%%%%%%%%%%%
\setlength{\extrarowheight}{0pt}
\begin{table}[t]
\centerline{
\begin{tabular}
{|c|c|c|c|c|c|c|}  \hline
classes: & (1)    & (123) &(132) &     (12)(34) \\ \hline
sizes:   & 1     &  4      &  4 & 3          \\  \hline
$\chi_1$ & 1  &   1  &   1   &  1     \\  %\hline
$\chi_2$ & 1   &   $\omega$       &  $\omega^*$    &    1          \\  %\hline
$\chi_3$ & 1  &   $\omega^*$  &  $\omega$ & 1          \\  %\hline
$\chi_4$ & 3  &   0  &  0  & -1         \\ \hline 
\end{tabular} }
\caption{\sl The character table of $A_4$, the pure  rotation subgroup of  
tetrahedron. 
The full point group symmetry  of the lattice 
action on  $A_3^*$ lattice is $S_4 \ltimes Z_2$}
\label{tab:A4}
\end{table}
%%%%%%%%%%%%%%%%%%%%%%%%%%%%%%%%%%%
%%%%%%%%%%%%%%%%%%%%%%%%%%%%%%%%%%

As in the discussion of $A_5$ symmetry group, there is an analogous 
four times four matrix $\CE$ which  splits all the $O(g) \in A_4$ into
block-diagonal form.  This matrix is  used to identify the irreducible
representations of the $A_4$ group with the ones of the twisted rotation
group $SO(3)'$.     Thus, in the  
continuum of $A_3^*$ lattice, we identify  $ \CE_{m\mu } z^{m}=z^{\mu}, \;
\CE_{ m4} z^{m}= \epsilon_{\mu \nu \rho } z_{\mu\nu \rho }/6 $  for the
 fields associated with links. Similarly, the two index fermions of the 
$A_3^*$ lattice are identified with the   continuum  fermions as 
$\xi_{mn} \CE_{m\mu}\CE_{n\nu}= \xi_{\mu\nu}$ and  
$\xi_{mn} \CE_{m\mu}\CE_{n4}= \epsilon_{\mu \nu \rho }\xi^{\nu \rho}/2$.
Further details, including a superfield formulation of the Blau-Thompson  
twist can easily be extracted from section four of ref.\cite{Kaplan:2005ta}.

\section {Two dimensional examples}

\subsection{A new twist of the  $\CN=(2,2)$  SYM theory}
 The  $\CN=(2,2)$  SYM  in $d=2$ can be  obtained by dimensional reduction
of  four dimensional  $\CN=1$ SYM theory down to two dimensions. 
The theory  possess a 
global $G=SO(2)_E \times SO(2)_{R_1} \times U(1)_{R_2}$ symmetry where 
$SO(2)_E$ is Euclidean Lorentz symmetry, $SO(2)_{R_1}$ is the symmetry due to  
reduced dimensions and  $U(1)_{R_2}$ is the $R$-symmetry of the theory prior 
to reduction. The twisted Lorentz group  $SO(2)'$ is the diagonal 
subgroup of $SO(2)_E \times SO(2)_{R_1}$.
 The vector $V_{\mu}$ transforming as $(2,1)_0$ and the scalar 
 $S_{\mu}$ transforming as  $(1,2)_0$ under $G$ 
 become  $(2)_0$ under $G'=SO(2)' \times U(1)_{R_2}$. 
We  complexify these  fields into   $z^{\mu}$ and   
$\mybar z_{\mu}$ as in  \Eq{vector}.
To see the transformation
 properties of the fermions is a little bit tricky, since the fermions 
transform under the  spin group of $SO(4)_E$, i.e, $SU(2)\times SU(2)$
(before the reduction). However,  
the reduction is inherently real, and   splits  $SO(4) \rightarrow  SO(2)_E
\times SO(2)_{R_1} $. 

In order to understand the transformation properties of fermions, we will
take advantage  of the relation between bispinors and vectors in four 
dimension. Let $v_{a}$,  $\mybar \omega$,and  $\omega$  be the gauge field,
the left and right handed spinors  of the  $d=4$  theory where $a=1, \ldots
4$. They transform under $SU(2) \times SU(2) \times U(1)_R$ respectively 
as $(2,2)_0  $,   $\;(2,1)_{-\frac{1}{2}}$, $\; (1,2)_{\frac{1}{2}}$.
We can turn the 
vector into a bispinor by using $\mybar \sigma_{a} = (1, i \vec \sigma)$ 
where  
$\vec \sigma$ is Pauli matrices and $1$  is the 
two dimensional identity matrix. 
$\omega_{\alpha} $  $(\mybar \omega_{\dot \alpha}) $  carries an undotted
  (dotted) spinor index $\alpha$ (  $\dot \alpha$) and the index structure of 
the  sigma matrix is  $(\mybar \sigma_{\mu})_{ \dot \alpha \alpha}$. Now, we
construct the bispinors  $v_{\dot \alpha \alpha } =
(v_{a}\mybar \sigma_{a})_{\dot \alpha \alpha }$ and   $ \mybar
\omega_{\dot \alpha} \omega_{\alpha}$. These two bispinor transform
identically under  $SU(2) \times SU(2) \times U(1)_R$ as $(2,2)_0$.
The $v_{\dot \alpha \alpha }$  can suitably be  expressed in terms of  
complexified  $SO(2)'$ doublets   $\mybar z_{\mu}$ and   $ z^{\mu}$.  
We have  
\beq
v_{\dot \alpha \alpha }= \sqrt 2 \left( \begin{array}{cc}
                                      \mybar z_1 & - z^2      \\         
                                        \mybar z_2 & \; \; z^1 
                                       \end{array} \right), {\rm \;\; and}  
\qquad 
 \mybar \omega_{\dot \alpha} \omega_{\alpha} 
=  \left( \begin{array}{cc}
 \mybar \omega_{\dot 1} \omega_{1}     & 
\; \mybar \omega_{\dot 1} \omega_{2}   \\         
 \mybar\omega_{\dot 2} \omega_{1} & 
\;  \mybar \omega_{\dot 2} \omega_{2} 
                                       \end{array} \right)
\eqn{bispin}
\eeq                 
where the columns are $SO(2)'$ doublets, ${\mybar z_\mu}$ and 
$\epsilon_{\mu \nu} z^{\nu}$. 
%Consider the  bispinor 
%$  \mybar \omega_{\dot \alpha} \omega_{\alpha}$. 
From \Eq{bispin}, 
 we see that  $\mybar \omega_{\dot \alpha}  \omega_{1}$ and 
 $\mybar \omega_{\dot \alpha}  \omega_{2}$  
has to be $SO(2)'$ doublets (vectors). 
Comparing with the columns of $v_{\dot \alpha \alpha }$ matrix,  we identify 
$\mybar \omega_{\dot \alpha}$ with an $SO(2)'$ vector,  $\omega_{ 1}$ with    
a scalar and  $\epsilon_{\mu \nu}  \omega_{ 2}$ with a second rank
antisymmetric tensor.  We label these accordingly as 
$\psi^{\mu}, \lambda, \xi_{\mu \nu}$.
Therefore, under the twisted  symmetry 
$SO(2)'\times U(1)_{R_2}$, we obtain the transformation 
properties of the fermions and bosons  as  
\beq
 \lambda \oplus  \psi^{\mu} \oplus \xi_{\mu \nu} 
\rightarrow 1_{\frac{1}{2}} \oplus 2_{-\frac{1}{2}} \oplus  
1_{\frac{1}{2}}, \qquad   z^{\mu} \oplus \mybar z_{\mu} \rightarrow  2_{0} +  2_{0}.
\eeq
This is indeed the two dimensional counterpart of the twist 
introduced by 
\cite{Marcus:1995mq, Blau:1996bx}.

The off-shell  supersymmetry transformation generated by the nilpotent 
scalar supercharge  is  given by 
\beq
&&Q \lambda= -id, \qquad Q d= 0, \cr
&&Q z^{\mu} = \sqrt 2 \psi^{\mu}, \qquad Q \psi^{\mu}= 0,   \cr
&&Q\mybar z_{\mu}=0,    \qquad  \mu=1,2  \cr
&&Q \xi_{\mu \nu}= -i \mybar \CF_{\mu \nu}, 
\eeq
where $d$ as usual is an auxiliary field introduced for the off-shell 
completion of the supersymmetry algebra $Q^2=0$. 
This is clearly a Blau-Thompson and Marcus  type twist, discussed in 
sections  \ref{sec:marcus} and \ref{sec:blau}. 
The action of the twisted theory is given by a $Q$-exact expression  
\beq
{\CL} &&= \frac{1}{g^2} Q \Tr  \Big[ \lambda( \half {id} +  
\half [\mybar {\cal D}_{\mu}, 
 {\cal D}^{\mu}])  + \fourthi \xi_{\mu\nu} {\CF}^{\mu\nu} 
\Big]   \cr
 &&= \frac{1}{g^2}  \Tr  \Big[ \eight ( [\mybar {\cal D}_{\mu} , 
 {\cal D}^{\mu}])^2 +   \fourth |{\CF}^{\mu\nu}|^2 + 
 \lambda  \mybar {\cal D}_{\mu} \psi^{\mu} +  
 \xi_{\mu\nu}  {\cal D}^{\mu} \psi^{\nu} \Big] \, .
\eqn{LT2}
\eeq
The $SO(2)' \times U(1)_{R_2}$ symmetry is manifest. 
Unlike the three and four dimensional counterparts, the action does
not have a $Q$-closed term and its  $Q$-invariance is manifest. This theory
can be made topological by regarding  $Q$ as a BRST. 
The study of the corresponding topological theory may be  interesting.

{\bf Square Lattice:\;} The two dimensional orbifold 
lattice action for $\CN=(2,2)$ theory 
yields the Blau-Thompson type twist in the continuum  \cite{Cohen:2003aa}.  
We observe that the 
fermions on the lattice are associated with  one site, two links and one 
face on each unit cell of the lattice.  
In the continuum, they fill, respectively,  the scalar, vector and second-rank 
antisymmetric tensor representation of $SO(2)'$.  The complex bosons are 
associated 
with the links (in both orientations) and they transform as vectors under 
$SO(2)'$.
The continuum  $U(1)_{R_2} $ symmetry of the twisted theory 
is an exact symmetry on the lattice.  

\subsection{The  $\CN=(4,4) $ SYM in $d=2$}
The $\CN=(4,4) $ SYM in $d=2$ can be obtained by dimensionally reducing the 
six dimensional  $\CN=1$ SYM theory down to two dimensions. The theory
possess  a $SO(2)_E \times (SU(2) \times SU(2))_{R_1} \times  SU(2)_{R_2}$  
symmetry group.  The $R_1$ symmetry is the internal symmetry due to reduction 
from six down to two dimensions and $R_2$ is the $R$-symmetry of  the theory 
prior to reduction.  
The  twisted theory 
 possesses  a   
$ SO(2)' \times U(1)_{R_1} \times SU(2)_{R_2}$ symmetry group. 
The orbifold lattice  only respects the $U(1)$ 
subgroup of the $SU(2)_{R_2}$ and therefore 
 we will express  the representations of the fields under 
$G' = SO(2)' \times U(1) \times U(1)$.  
The six bosonic fields transform under $G'$ as
\beq 
z \oplus \mybar z \rightarrow 1_{1,0} \oplus 
1_{-1,0},  \qquad
z^{\mu} \oplus \mybar z_{\mu} \rightarrow 2_{0,0} \oplus 2_{0,0}
\eeq 
The eight fermion spits into two groups of four as 
\beq 
&&\lambda \oplus  \psi^{\mu}  \oplus \psi_{\mu \nu}  \rightarrow  
1_{\frac{1}{2}, -\frac{1}{2}} \oplus 
2_{-\frac{1}{2}, \frac{1}{2}} 
\oplus 1_{\frac{1}{2}, -\frac{1}{2}}, \cr
&& \mybar \lambda  \oplus \mybar \psi_{\mu} \oplus \mybar 
\psi^{\mu \nu} \rightarrow  
 1_{\frac{1}{2}, \frac{1}{2}} \oplus
2_{-\frac{1}{2}, -\frac{1}{2}}
\oplus  1_{\frac{1}{2}, \frac{1}{2}} 
\eeq
This twist is 
examined in detail in \cite{Geyer:2001qy}. 

{\bf Square  Lattice:\;} The two dimensional orbifold 
lattice action for $\CN=(4,4)$ theory 
yields the Blau-Thompson type twist in the continuum.  Having twice as many
fermion with respect to  $\CN=(2,2)$ theory, each $p$-cell on the lattice
accommodates twice as many fermions (in  opposite orientation).  Besides
discrete lattice symmetries, the lattice also possess a continuous 
 $U(1) \times  U(1)$ symmetry.  
In the continuum, these symmetries enhances to $ SO(2)' \times U(1)_{R_1}
\times SU(2)_{R_2}$ symmetry of the twisted theory. The superfield formulation 
of the twisted continuum and lattice theory is given in \cite{Cohen:2003qw}. 

\subsection{The $\CN=(8,8) $ SYM in $d=2$}
The  $\CN=(8,8) $ SYM theory possess an $SO(2)_E \times  SO(8)_R$ symmetry 
group.  The global symmetry of the twisted theory is  $G'= SO(2)' \times 
SU(2) \times SU(2) \times U(1)$. The ten bosons transform  under $G'$ as
\beq 
&&z^{\mu} \oplus \mybar z_{\mu} \rightarrow (2,1,1)_0 \oplus (2,1,1)_{0} \cr
&&z_{\mu \nu} \oplus \mybar z^{\mu \nu} \rightarrow (1,1,1)_1 \oplus 
(1,1,1)_{-1}, \qquad   
\tilde z  \rightarrow  (1,2,2)_0
\eeq 
For the sixteen fermionic degree  of freedom,  we obtain 
\beq 
&& \lambda \oplus \psi^{\mu} \oplus \psi_{\mu \nu}  \rightarrow 
(1,2,1)_{\frac{1}{2}} \oplus  (2,2,1)_{-\frac{1}{2}} \oplus 
(1,2,1)_{\frac{1}{2}} \cr
&&  \mybar \lambda \oplus  \mybar \psi^{\mu}  \oplus
  \mybar \psi^{\mu \nu}  \rightarrow 
(1,1,2)_{-\frac{1}{2}}  \oplus  (2,1,2)_{\frac{1}{2}}  \oplus  
(1,1,2)_{-\frac{1}{2}}
\eeq 

{\bf Square  Lattice:}  The fermions are distributed in multiples of four 
to each $p$-cell as $4(1\oplus2 \oplus 1)$.  
Four of the bosons (labeled as $\widetilde z$) are 
associated with site,  four of them ($z^{\mu}$ and $\mybar z_{\mu}$)
 with  the links each accommodating two,  
and two of them $z_{\mu\nu}$ and $\mybar z^{\mu\nu}$ on the face diagonal. 
For details,  see  \cite{Kaplan:2005ta}. 

{${\pmb A_2^{*}}$ {\bf (Hexagonal) Lattice:} }
The $A_2^*$ orbifold  lattice action possess a point group symmetry 
 $S_3 \ltimes Z_2$, 
where $S_3$ is  the permutations  of the chiral multiplets and  $Z_2$ is 
the  inversion  symmetry  swapping  chiral and antichiral multiplets. Another
discrete symmetry of the action is charge conjugation.    
Following the analysis of the $A_3^{*}$ and $A_4^{*}$ lattices, it is
sufficient   to construct the pure rotation subgroup $A_3$ of $S_3$ to make
connection to the twisted form. However, 
$A_3$ is  an  abelian cyclic group and it only possess
one dimensional representations.  This is not a problem. Recall that 
the two dimensional vector representation of $SO(2)'$  is also reducible when
we regard it in its spin group, $U(1)'$. 
The $A_3$ character table has two complex conjugate  characters 
$ \chi_2 = (1, e^{2\pi i/3}, e^{-2\pi i/3})$ and  $ \chi_3 = (1, 
e^{-2\pi i/3}, e^{2\pi i/3})$. 
 These two complex conjugate
representation of the $A_3$ group has to be regarded as one two dimensional 
representation. 
 The sum   $ \chi= \chi_2 +  \chi_3 = (2, -1,-1)$ is a two dimensional real
 character  and is  irreducible over $\mathbb R$.
Alternatively, we can also work with the full nonabelian point group 
symmetry of the lattice. The   $S_3 \ltimes Z_2$ group has two dimensional 
representations and a little bit more information than we need here. 
\footnote{The same lattice structure also emerges for the spatial lattice of 
$\CN=8$ theory in $d=3$ dimensions. See the footnote.\ref{foot:ham}}

As in the $A_3^{*}$ and $A_4^{*}$  lattices,   there is a $A_3$-invariant  
subspace of the link fields, and consequently,   
the link field  splits into a singlet and a  two dimensional representation.  
We obtain the characters  as 
\beq
&&\chi(z^m)= \chi( \mybar  z_m) = (3, 0,0) = \chi \oplus
 \chi_1 \cr
&&\chi(z)=  \chi(\lambda)= \chi(\mybar \lambda) = (1,1,1) =    \chi_1   
\eeq
Therefore, the sixteen fermions and ten bosons splits as  
\beq
&&{\rm fermions} \rightarrow 4 ( \chi_1 \oplus \chi
  \oplus \chi_1 )\\  
&&{\rm bosons} \rightarrow 4  \chi_1 \oplus 2 \chi_3 \oplus 2 \chi_1 
\eeq
as  in the continuum twisted theory  discussed above. 
There is also an analogous  matrix $\CE$  
which  maps the irreducible
representations of $S_3$ (or $A_3$) into the ones of twisted rotation 
group  $SO(2)'$. The matrix  $\CE$   brings the 
group actions of $A_3$  into a block diagonal 
form. 
Thus, in the 
continuum of $A_2^*$ lattice, the fields associated with the links  become 
vector and scalar representation of $SO(2)'$. Explicitly. we have 
$ \CE_{m\mu } z^{m}=z^{\mu}, \;
\CE_{ m3} z^{m}= \epsilon_{\mu \nu } z_{\mu\nu  }/2 $ and similar mappings 
for other fields. 
\section{Conclusions and prospects}
Certainly  one of the most 
bizarre features of the orbifold lattices was associating 
spinless bosons of the continuum theory with the link fields which 
 transform
nontrivially on the lattice, and  associating double valued 
 spinor representation of the continuum with the single valued 
 representations of the point group of the  lattice \cite{Kaplan:2002wv,
Cohen:2003aa, Cohen:2003qw, Kaplan:2005ta, Kaplan:2003uh}.
Remarkably, the orbifold
 lattice in the continuum gave Lorentz invariant, highly supersymmetric 
 theories with no or little fine tuning. This work  hopefully demystifies 
the orbifold lattices
 by relating  them  to the twisted versions of  
supersymmetric theories. Many  twisted theories 
arise, in the continuum,  as a courtesy of the orbifold projection. 
These twisted
versions  are often worked in the context of topological field theory, and 
we hope this work leads to further, fruitful  interplay between these 
two branches.
Before  moving to the prospects, let us give  the summary of our  results:
\begin{itemize}
{\item   The orbifold lattices, in the continuum, reproduce the  
 Marcus and Blau-Thompson  twists  of the extended supersymmetric theories. 
Conversely, it is
possible to discretize (with a well-defined recipe)  
the Marcus and Blau-Thompson twists of 
the extended supersymmetric theories  to obtain the  orbifold lattice action. }
{\item  The point group symmetry of the orbifold lattice is a subgroup of 
the  twisted  Lorentz group, and not the  real Lorentz group. }
{\item  The exact supersymmetries on the orbifold lattices are the 
nilpotent spin zero, scalar supersymmetries of the continuum twisted theory.} 
{\item  The $p$-form fields on the continuum are naturally associated 
with $p$-cells on the hypercubic lattices.  For more symmetric 
$A_d^*$ lattices, the irreducible representations of lattice rotation group 
are in one to one  correspondence with the representations of twisted
rotation  group.}
\end{itemize}

It is also possible to understand the spatial orbifold lattices 
\cite{Kaplan:2002wv} and 
 deconstruction of  higher dimensional supersymmetric  theories 
\cite{ArkaniHamed:2001ie} from the viewpoint of the present work. 
They correspond to latticization of 
partial or half twisted versions  of the corresponding 
target field theories. Also, a few new  partial 
twisting  of  $\CN=2$, and $\CN=4$   in $d=4$
supersymmetric Yang-Mills theory seems to exist.   

It is clear that the twisted versions of the supersymmetric theories are in
a more peaceful existence with lattice. The main point is that in the 
twisted theories some of the supercharges are spin zero scalars,   and they 
do not make any reference to the underlying structure of spacetime.  
Even when carried into the lattice,  the supersymmetry algebra  $Q^2=0$ 
still holds with no reference to finite  lattice translations.
We believe this relation is the 
key for the lattice regularization for a larger class of supersymmetric
theories.  
It seems that  twisted versions of certain sigma-models 
in two dimensions may provide good  opportunities. Some theories of this 
type are known to have an isolated, discrete vacua, 
a  discrete spectrum and mass gap. 

There are also  interesting directions to explore in the continuum 
twisted versions . An interesting class of theories 
arises from the  $SO(4)' \times U(1)$  and  $\CQ=1$ symmetry preserving 
deformations of  the  twisted action \Eq{LS4}.  Clearly, there is a 
few  parameter family of deformations of  \Eq{LS4} satisfying 
these requirements. For example,  
altering  $\CL_2$ into  $ { \CL}_{2}= 
Q  \Tr  \Big( 
c_1 \fourthi \xi_{\mu\nu} \CF^{\mu\nu} + c_2 \textstyle{\frac{1}{12 \sqrt2}}
\xi^{\nu \rho \sigma} {\cal D}^{\mu} z_{\mu \nu \rho \sigma} 
\Big) $, where $c_1$ and $c_2$ are deformation parameters   is  of this type. 
For $\CL_3$, the two $SO(4)' \times U(1)$ singlets are glued to each other 
because of $\CQ=1$ supersymmetry, however an overall parameter  is possible.  
Only for a special choice of the deformation  parameters (for example 
$c_1=c_2 =1$ etc), and    in  flat spacetime, 
this  theory  \Eq{LS4} is  a rewriting of  $\CN=4$ SYM and is under the 
strong protection of underlying  higher symmetry, sixteen supersymmetries.  
The other theories, for example  with $c_1\neq c_2$ may be 
worth exploring, both in flat and curved  spacetimes.     
The most natural
framework to think about such deformations seems to be (Euclidean) D3-branes 
wrapped on  curved four manifolds. It is well-known that  the world-volume of 
the wrapped  D-branes do not realize the usual form of the supersymmetry, 
but a twisted version  of it. 
The constructions in this paper can be considered as   
a straightforward realization of this idea, because underlying 
manifold (in continuum) is flat, $d$-dimensional torus $T^d$.   

Another issue which arises from the twisted versions  are related to 
BPS solitons.  As in the Witten's treatment of Donaldson theory 
\cite{Witten:1988ze}, where
instantons appears as fixed points of supersymmetry transformations, 
the vanishing of fields under  $\widetilde Q$ in 
\Eq{Qtwisted} produce a complex generalization  of the instanton equation. 
Similar considerations  for the Blau-Thompson 
twists of  $d=3$,   $\CN=8$ theory  yields a complex generalization of 
monopole  equation. It is desirable to understand these solitons  
in more detail, and 
in particular the  structure of their  moduli spaces.  Research in this
direction is ongoing.  
 
\acknowledgments
I would like to thank David B. Kaplan for teaching me the r\^ole of  
representation theory in orbifold lattices. This led me to think about 
the  relation to the twisted theories. I am grateful for many useful 
discussions about
this work with  Simon
Catterall, David B. Kaplan and Takemichi Okui. 
I also thank  to Michael 
Douglas, Joel Giedt,   Ami Hanany,  Andreas Karch, Ami Katz,  Adam Martin,
  Scott Thomas for conversations. 
This work was supported by DOE grant  DE-FG02-91ER40676.

\appendix
\section{Twistings by discrete $R$-symmetries and  finite spin groups}

In this appendix, I will briefly sketch an alternative view on twisting, from
the viewpoint of discrete groups. For clarity,  I will distinguish the
groups with double valued representation from the ones with single valued 
representations. For example, 
spin groups will be treated differently from the orthogonal group. 

 In this paper, we considered theories with 
sufficiently large $R$-symmetries such that a nontrivial homomorphism from the 
full Lorentz group to the $R$-symmetry group was possible. We
performed twists of a rather simple kind by  constructing the diagonal sum 
\beq
 {\rm Diag}(Spin(d) \times Spin(d))= Spin(d)'
\eeq
 At the end, only
integer spin representations appeared in $Spin(d)'$. These representations 
 are $p$-form fields  and are  the   shared representations with $SO(d)'$.
That means, in the twisted theory, we really do not need to think of spin 
group anymore since there are no spinor representations at all. One of the 
 main observation of this paper is that the point group symmetry of the 
 supersymmetric  orbifold lattices is a finite subgroup of the $SO(d)'$. 
 
Can we understand  the above construction in the language of finite groups? 
The answer is positive and complementary to the approach in the bulk of this 
paper. The answer clearly requires the knowledge of finite subgroups of spin
groups. The classification of these groups is well know,  and 
these  are the spin groups of the point group symmetries.  Let us
consider a particular case:  The spin group of $SO(3)$  is 
$Spin(3)=SU(2)$. It is related to $SO(3)$ by a two to
one map  $SU(2)/\pm1 =SO(3)$. Let us call this map $\pi$. We have 
$\pi: SU(2) \rightarrow SO(3)$. Given any  finite subgroup $G_{\rm f}$ of 
$SO(3)$, we
can look for  $\pi^{-1}(G_{\rm f})$.  This gives a list of finite subgroups 
of the $SU(2)$, which we label as  $\widetilde G_{\rm f}$. Examples of
$\widetilde G_{\rm f}$ are 
 $\widetilde A_3,  \widetilde A_4, \widetilde S_3,  
\widetilde S_4$.  These are respectively, the spin groups (doubling) 
 of the finite groups   $  A_3, A_4,  S_3,   S_4$  which  frequently appeared 
as the point group symmetry of the lattices. The doubled-groups  
$\widetilde G_{\rm f}$ admit spinor representations. 
 The number of conjugacy classes (hence characters)  of   
$\widetilde G_{\rm f}$ is always larger then the one of $ G_{\rm f}$, but
usually not twice as much. 

Let us consider a finite  subgroup of $Spin(d)_L \times Spin(d)_R$, which
we will label as  $\widetilde G_{\rm f} \times \widetilde G_{\rm f}$. The
first one of these corresponds  to spacetime and the latter corresponds to 
$R$-symmetry.  Let us assume the spacetime is discretized.  Then 
the fields transforming in irreducible representations of  $Spin(d)$ will 
split into irreducible representations of  $\widetilde G_{\rm f}$.  
For  low dimensional representation of $Spin(d)$, there is usually a single 
corresponding  representation in $\widetilde G_{\rm f}$ and there is no
splitting. Of course, high dimensional representations of the $Spin(d)$ 
will split into many representation of  $\widetilde G_{\rm f}$ since, simply, 
the representations of  $\widetilde G_{\rm f}$ are finite dimensional. This 
is similar to the level spitting of an atom inserted into a field of crystal 
potential  which has a finite symmetry group.  Assuming the 
perturbing potential has a  lower symmetry,  
the degeneracies are determined by the representations of the  perturbation.
In our case,  for the fields appearing in Lagrangian, there is usually
just a single  representation to be matched with in lattice. 
In order to obtain the 
orbifold lattices, it seems inevitable that the internal $R$-symmetry 
 has to be  restricted to a finite spin group as well.  This finite $R$-spin
 group has to be  necessarily identical to the  $\widetilde G_{\rm f}$ 
of spacetime for the desired outcome. 

Let us  consider an example: A spacetime spinor  fermion  
$\psi_{\dot \alpha, \alpha}$
in the bi-spinor  representation of 
$Spin(3)_L \times Spin(3)_R$.  It transforms under 
$Spin(3)_L \times Spin(3)_R$  as $\psi \rightarrow L\psi R^{\dagger}$ with
 obvious action of $L$ and $R$. 
Let us assume that  $L$  is  an element of double-group 
$\widetilde G_{\rm f}$, and let us
consider a particular combination of the field such as $\Tr \psi$. (I will
come back to other components momentarily.)  
Then, it is clear that whatever $L$ action we choose, the field  $\Tr \psi$
will remain invariant as long as I restrict $R$ to discrete operations 
$R=L$. In that case  $\Tr \psi  \rightarrow \Tr L \psi R^{\dagger} = 
\Tr L \psi L^{\dagger}=\Tr \psi .$  
That means the field  $\Tr \psi$ is invariant 
under the diagonal sum of  $\widetilde G_{\rm f} \times \widetilde G_{\rm
  f}$. Let us call this diagonal subgroup   $\widetilde G_{\rm f}'$. 
Then we can define the twisted discrete point group as
\beq
  {\rm Diag}(\widetilde G_{\rm f} \times 
\widetilde G_{\rm f}) =\widetilde G_{\rm f}' \; .
\eeq
As in the case of its continuous counterpart, there are no double-valued
representations appearing in  $\widetilde G_{\rm f}'$. Therefore, the group
is really just  $G_{\rm f}'$, which is a subgroup of the twisted rotation
group $SO(d)'$. 

Now, let us come back to  the other components of the bispinor field and
treat them slightly  more rigorously.  
The spin group $\widetilde A_4$ has seven conjugacy
classes (see, for example, \cite{Landau}, page 393)  whereas as shown in 
Table.\ref{tab:A4}, the $A_4$ has only
four.  Rather than  examining the details of representation of 
$\widetilde A_4$, 
we  want to use necessary information to see 
 the fate of the other components  of   
$\psi_{\dot \alpha, \alpha}$ field. 
The conjugacy classes with their multiplicities
are $$(1),\; S, \; 4(123),
\;4(132), \;4(123)S, \;4(132)S, \; [ 3(12)(34)+ 3(12)(34)S]$$ 
where $S$ is the $2\pi$  rotation such that $S^2= 1$. The character for the
two dimensional spinor representation is 
$\chi(\psi_{\dot \alpha})= (2,-2, 1,-1, -1,1,0)$.  Under the
diagonal  $\widetilde A_4^{'}$, $\psi_{\dot \alpha, \alpha} $ transform as 
$\chi(\psi_{\dot \alpha, \alpha})= \chi(\psi_{\dot \alpha}) \times 
\chi(\psi_{\dot \alpha})  = 
(4,4,1,1,1,1,0)$.  The product splits into two representations, and these 
are indeed common  representations with  $A_4^{'}$. Therefore, 
it is sufficient to 
 inspect the character table of $A_4$. We conclude 
$\chi(\psi_{\dot \alpha, \alpha})= \chi_4 \oplus \chi_1$ as we expect. The
bispinor 
$\psi_{\dot \alpha, \alpha}$ splits into single valued, a tree dimensional  
vector representation and a one dimensional  scalar  representation under 
 $A_4'$. Explicitly, we have 
\beq
\psi_{\dot \alpha, \alpha} =  
(\psi^0 1_2 + \psi_{\mu} \sigma^{\mu})_{\dot \alpha, \alpha},  \qquad 
{\rm or} \;\; 
\psi^0 = \half \Tr \psi, \;\;   \psi_{\mu}= 
\half \Tr \psi  \sigma_{\mu}
\eeq
 where $\sigma_{\mu}$ are the usual Pauli
matrices. 

Finally, I do not know a lattice formulation 
 which is supersymmetric and invariant 
under $\widetilde G_{\rm f} \times 
\widetilde G_{\rm f}$ or $\widetilde G_{\rm f} \times$ (full $R$-symmetry). 
The difficulty is that;  under 
the real spacetime symmetry group  scalars, gauge bosons
and fermions are treated on very different  footing on the  lattice. However,
the twisted version  happily accommodates all while preserving a subset of 
supersymmetry.   

\bibliography{nc}
\bibliographystyle{JHEP} 
\end{document}